\documentclass[fleqn,usenatbib]{mnras}
\usepackage{graphicx}
\usepackage{amsmath}
\usepackage{amssymb}
\graphicspath{{./image/}}
\usepackage{lmodern}
\usepackage{hyperref}
\usepackage{makecell}

\newcommand{\HI}{\rm H~{\sc i }}
\newcommand{\HII}{\rm H~{\sc ii }}
\newcommand{\HeI}{\rm He~{\sc i }}
\newcommand{\HeII}{\rm He~{\sc ii }}

\newcommand{\TB}{\delta T_{\rm b}}
\newcommand{\MSUN}{{\rm M}_{\odot}}

\newcommand{\XHII}{x_{\rm HII}}
\newcommand{\TS}{T_{\rm S}}
\newcommand{\TK}{T_{\rm K}}
\newcommand{\TCMB}{T_{\gamma}}
\newcommand{\lya}{\rm {Ly{\alpha}}}
\newcommand{\OmegaB}{\Omega_{\rm B}}
\newcommand{\Omegam}{\Omega_{\rm m}}

\title[Composite X-ray binary spectra with MAXI]{Cosmological implications of the composite spectra of galactic X-ray binaries constructed using MAXI data}

\author[Islam et.al]{
\parbox[t]{\textwidth}{
Nazma Islam$^1$\thanks{Email: nazma.syeda@cfa.harvard.edu},
Raghunath Ghara$^2$\thanks{Email: ghara.raghunath@gmail.com},
Biswajit Paul$^3$, T. Roy Choudhury$^4$ and Biman B. Nath$^3$} 
\vspace*{6pt} \\
$^1$ Center for Astrophysics $\mid$ Harvard $\&$ Smithsonian, 60 Garden Street, Cambridge, MA 02138, USA\\
$^2$ Department of Astronomy \& Oskar Klein Centre, AlbaNova, Stockholm University, SE-106 91 Stockholm, Sweden\\
$^3$ Raman Research Institute, Sadashivanagar, Bangalore-560080, India\\
$^4$ National Centre for Radio Astrophysics, TIFR, Post Bag 3, Ganeshkhind, Pune 411007, India\\}

\date{Accepted ?; Received ??; in original form ???}

\pubyear{?}

\date{}

\begin{document}
\label{firstpage}
\pagerange{\pageref{firstpage}--\pageref{lastpage}}
\maketitle

\begin{abstract}
We have investigated the long term average spectral properties of galactic X-ray binaries in the energy range of 3--20 keV, using long term monitoring data from {\it MAXI}--Gas Slit Camera (GSC). These long term average spectra are used to construct separately the composite spectra of galactic High Mass X-ray binaries (HMXBs) and Low Mass X-ray binaries (LMXBs). These composite spectra can be described empirically with piece-wise power-law with three components. X-rays from HMXBs are considered as important contributors to heating and ionization of neutral hydrogen in the intergalactic medium during the Epoch of Reionization. Using the above empirical form of the composite HMXB spectra extrapolated to lower energies as an input, we have studied the impact of these sources on the 21-cm signal using the outputs of $N$-body simulation and 1D radiative transfer. The heating due to the composite spectrum is less patchy compared to power-law spectrum with a spectral index $\alpha = 1.5$, used in previous studies. The amplitude of the heating peak of large scale power spectrum, when plotted as a function of the redshift, is less for the composite spectrum. 
\end{abstract}


%%%%%%%%%%%%%%%%%%%%%%%%%%%
%%%%%%%%%%%%%%%%%%%%%%%%%%%
\section{Introduction}
\label{sec:intro}
X-ray binaries are gravitationally bound binary systems, consisting of a compact object like a neutron star or a black hole, together with a main sequence or a supergiant as the companion star. Accretion, in which matter from the companion is captured by the compact object either from the stellar wind or through Roche lobe overflow, is the main source of energy for these systems. Depending on the mass of companion star (M$_{\rm C}$), they are broadly classified as High Mass X-ray binaries (M$_{\rm C} \geq$ 10 M$_{\odot}$) and Low Mass X-ray binaries (M$_{\rm C} \leq$ 1 M$_{\odot}$). HMXBs are further broadly classified into Be-HMXBs and Supergiant HMXBs depending on the type of the companion star. LMXBs are broadly classified into Z and Atoll sources, based on the patterns traced in their X-ray colors diagram \citep{hasinger1989}.

Since the discovery of Sco X-1 \citep{giacconi1962}, X-ray binaries have been extensively studied in various energy bands. More than half a century of investigations of these sources with various X-ray observatories show a diverse range of spectral and temporal characteristics. The spectral and temporal properties of X-ray binaries are related to the system parameters (spin period or orbital period) and accretion flows around the compact object (see \citealt{white1983,nagase1989,paul2017} -- for description of properties of accreting pulsars; \citealt{remillard2006} - for description of properties of black hole binaries). These X-ray binaries are highly variable and show large intensity and spectral variations. Most of the spectral and timing information about these X-ray binaries come from the time limited window of pointed observations with observatories such as {\it Ginga}, {\it ASCA}, {\it RXTE}--PCA, {\it Chandra}, {\it XMM--Newton}, {\it Suzaku} etc. These observations are carried out in the selected states of X-ray binaries such as during outbursts, transition between different spectral states and very few quiescent states. There is a substantial difference between the spectra of X-ray binaries in its different intensity states. For example, the spectra of Be-HMXBs show a strong Comptonisation component during its outbursts, whereas in quiescence, they have either power-law spectra or faint thermal spectra \citep{tsygankov2017}. BH binaries show various spectral states, with either a strong Comptonisation component or a disk dominated thermal component \citep{remillard2006}. The NS LMXBs in quiescence, show a presence of a non-thermal power-law component along with a soft thermal component, whereas, in outbursts, they have a strong disk dominated emission with a hard power-law \citep{fridriksson2011,lin2007}.

On the other hand, X-ray telescopes {\it Chandra}, {\it XMM--Newton}, and {\it NuSTAR} have identified a population of X-ray sources in external galaxies, many of them as X-ray binaries \citep{fabbiano2006,zhang2012,vulic2018}. Previous studies by \citet{mineo2012, pacucci2014} constructed the spectral energy distribution of High Mass X-ray binaries in energy range 0.5--8.0 keV, using {\it Chandra} observations of nearby galaxies. These observations of extra-galactic X-ray binaries are essentially tens of kiloseconds `snapshot' observations, which includes a large number of sources, in different spectral states, thus providing an effective average spectrum. However, our work focuses on investigating the average spectra of galactic X-ray binaries using their long term monitoring data.

Several X-ray all sky monitors have been continuously monitoring X-ray binaries for the past few decades. Although the main objective of X-ray all sky monitors is to look for new X-ray transients, their long and uniform coverage of the X-ray sky has led to many orbital and super-orbital modulation studies (\citealt{wen2006} with {\it RXTE}--ASM and \citealt{corbet2007,corbet2013} with {\it Swift}--BAT). However, the long term average spectral properties of X-ray binaries have not been studied before, partly due to the lack of an all sky monitor with good spectral capabilities. The wide field monitoring of the whole sky with the Monitor of All Sky X-ray Image (MAXI), along with its good spectral capabilities, have now made it possible to investigate the true long term averaged spectral properties of X-ray binaries. 

%%%%%%%%%%%%%%%%%%%%%%%%%%%%

The HMXBs are often considered as important contributors to the heating and reionization of very high redshift intergalactic medium (IGM) during the Cosmic Dawn and the `Epoch of Reionization' (EoR) \citep{power2009,mirabel2011,power2013, fialkov2014, pacucci2014}. X-ray photons emitted from accreting X-ray binaries have longer mean free path than the stellar UV photons and are also capable of producing several secondary ionization, thereby enhancing the ionizing efficiency. However, the details of the physical processes during the Cosmic Dawn and EoR are unknown. 

The observations of the redshifted 21-cm signal emitted by the neutral hydrogen in the IGM can provide us with the various details about the states of IGM, properties of early sources etc. during these epochs. Thus, several experiments have been designed in order to detect the signal from such high redshifts. Experiments such as EDGES \citep{2010Natur.468..796B}, EDGES2 \citep{monsalve2017,EDGES2018}, SARAS \citep{2015ApJ...801..138P},  SARAS2 \citep{singh2017}, BigHorns \citep{2015PASA...32....4S}, SciHi  \citep{2014ApJ...782L...9V}, LEDA \citep{price2018} etc. aim to detect the average 21-cm signal signal. On the other hand, several radio telescopes such as  the Low Frequency Array (LOFAR)\footnote{http://www.lofar.org/} \citep{2017ApJ...838...65P}, the Precision Array for Probing the Epoch of Reionization (PAPER)\footnote{http://eor.berkeley.edu/} \citep{parsons13}, the Murchison Widefield Array (MWA)\footnote{http://www.mwatelescope.org/} \citep{bowman13}, the Hydrogen Epoch of Reionization Array (HERA) \citep{2017PASP..129d5001D}  etc. have dedicated their resources to detect the signal in terms of statistical quantities such as the power spectrum etc.

Various cosmological simulations predicting the redshifted 21-cm signal from the Cosmic Dawn and EoR, use an input X-ray spectra to model the contribution of X-ray photons produced from HMXBs \citep{pritchard2007,power2013,fialkov2014}. However, there are disagreements related to the actual model of the input HMXB spectra used in various simulations. \cite{pritchard2007} and \cite{power2009} used an empirical power-law with energy-index $\alpha \sim$ 1, whereas \cite{mirabel2011} and \cite{power2013} used the spectra of Cyg X--1, a Galactic black hole HMXB, as the template for HMXB spectra. \cite{fialkov2014}  and \cite{das2017} used a X-ray binary spectral energy distribution calculated from a XRBs population synthesis simulation by \cite{fragos2013}, which is motivated by {\it RXTE}--PCA observations of galactic neutron star and black hole binaries. \cite{pacucci2014} used the 0.5--8.0 keV SED constructed from Chandra observations of starburst galaxies by \cite{mineo2012} as input X-ray spectra. 

These fiducial HMXB spectra, however, do not represent the true average spectral energy distribution of HMXBs for the following reasons:
\begin{itemize}
 \item As mentioned previously, {\it RXTE}--PCA observations of galactic NS and BH binaries do not represent the long term average spectra of X-ray binaries because these are  pointed observations carried out during very selected states of XRBs.
 \item Using Cyg X--1 spectra as an input template for HMXB spectra is erroneous because it does not represent the spectral behaviour of the general HMXB population, which is dominated by accreting X-ray pulsars. 
 \item Studies of normal galaxies with NuSTAR show that their X-ray spectra fall rapidly above 10 keV \citep{lehmer2015,yukita2016}, whereas X-ray binaries typically have a harder spectrum.
\end{itemize}

Long term observations of galactic X-ray binaries with MAXI makes it possible to construct composite X-ray spectra of galactic X-ray binaries. 
In this work, we have investigated the long term average spectral properties of galactic X-ray binaries using {\it MAXI}--GSC data and constructed the composite galactic X-ray binary spectra, separately for High Mass X-ray binaries and Low Mass X-ray binaries. The composite HMXB spectra is then used in estimating the contribution of X-ray binary heating to the 21-cm signal from the Cosmic Dawn and the Epoch of Reionization.

%%%%%%%%%%%%%%%%%%%%%%
\section{Observations and Data Analysis}
\label{sec:obsdata}
\subsection{Monitor of All Sky X-ray Image}
{\it MAXI} is an X-ray all sky monitor operating on the International Space Station (ISS) since 2009 \citep{matsuoka2009}. {\it MAXI}--GSC has the highest sensitivity and energy resolution among past and currently active all sky monitors: it reaches a 1 day sensitivity of 9 mCrab (3 $\sigma$), compared to 15 mCrab for {\it RXTE}--ASM and 16 mCrab for {\it Swift}--BAT \citep{krimm2013}. The main instrument onboard is the Gas Slit Camera, which operates in the energy band 2--30 keV \citep{mihara2011}. GSC consists of six units of large area position sensitive Xenon proportional counters. It has 85\% coverage of the entire sky in every 92 minutes of the ISS orbit. It has an energy resolution of 16\% (FWHM) at 6 keV. The typical daily exposure of GSC is 4000 cm$^{2}$s and the daily 5$\sigma$ source sensitivity $\sim$ 15 mCrab \citep{sugizaki2011}.
\par
We have extracted the spectra of 57 X-ray binaries from Modified Julian Day: 55058 to 56970 ($\sim$ 5.2 years), which have enough statistics to allow spectral fitting, using the MAXI on-demand data processing \footnote{http://maxi.riken.jp/mxondem/} \citep{nakahira2012}. Since X-ray binaries are highly variable, the sample of X-ray binaries are chosen such that its average X-ray flux during the 5.2 years of {\it MAXI}--GSC at least 1.8 $\times 10^{-10}$ ergs cm$^{-2}$ s$^{-1}$. For some bright X-ray binaries such as Sco X--1, GRS 1915+105, Cyg X--1, GX 17+2, GX 349+2 and GX 5--1, small instrument calibration uncertainties can be seen in the residuals to the best fitted spectrum, in Figure \ref{hmxb_spectra} and \ref{lmxb_spectra}. We have added 2 \% systematic errors in the energy band 4--6 keV in the spectra of these sources. 
\par
There are 17 High Mass X-ray binaries and 40 Low Mass X-ray binaries in the sample, given in Table 1 with their average X-ray fluxes and luminosities. The distances to the X-ray binaries are taken from \citet{islam2016}.

\begin{table*}
\centering
\caption{List of galactic X-ray binaries considered in this paper to construct the composite spectrum of the X-ray binaries, along with its average X-ray fluxes and luminosities in 3--20 keV energy band. The distance to the X-ray binaries are taken from \citet{islam2016}.}
\begin{tabular}{| c c c c c |}
\hline
Source          & Type                                         & Distance(kpc)  & Average Flux (ergs cm$^{-2}$ s$^{-1}$) & Average Luminosity (ergs s$^{-1}$)\\
                &                                              &                & in 3--20 keV                           & in 3--20 keV \\
\hline
Cyg X--1        & HMXB (BH)                                    & 1.83           & 1.08 $\times 10^{-8}$                  & 4.3 $\times 10^{36}$ \\
Cyg X--3        & HMXB (BHC)                                   & 7.2            & 5.09 $\times 10^{-9}$                  & 3.2 $\times 10^{37}$ \\
Vela X--1       & sgHMXB pulsar                                & 2.2            & 3.5 $\times 10^{-9}$                   & 2.0 $\times 10^{36}$ \\
Cen X--3        & sgHMXB pulsar                                & 9.0            & 2.7 $\times 10^{-9}$                   & 2.6 $\times 10^{37}$ \\
GX 301--2       & Be-HMXB pulsar                               & 3.1            & 2.3 $\times 10^{-9}$                   & 2.6 $\times 10^{36}$ \\
4U 1700--37     & HMXB                                         & 1.9            & 2.2 $\times 10^{-9}$                   & 9.5 $\times 10^{35}$\\
A 0535+262      & Be-HMXB pulsar                               & 3.8            & 1.7 $\times 10^{-9}$                   & 2.9 $\times 10^{36}$\\ 
GX 304-1        & Be-HMXB pulsar                               & 1.3            & 1.1 $\times 10^{-9}$                   & 2.2 $\times 10^{35}$\\ 
X Per           & Be-HMXB pulsar                               & 0.8            & 9.9 $\times 10^{-10}$                  & 7.6 $\times 10^{34}$\\
OAO 1657--415   & sgHMXB pulsar                                & 7.1            & 9.3 $\times 10^{-10}$                  & 5.6 $\times 10^{36}$\\
4U 1538--52     & sgHMXB pulsar                                & 4.5            & 5.0 $\times 10^{-10}$                  & 1.2 $\times 10^{36}$\\ 
EXO 2030+375    & Be-HMXB pulsar                               & 3.1            & 4.1 $\times 10^{-10}$                  & 4.7 $\times 10^{35}$\\ 
GRO J1008--57   & Be-HMXB pulsar                               & 5.0            & 4.0 $\times 10^{-10}$                  & 1.2 $\times 10^{36}$ \\
3A 1145--616    & Be-HMXB pulsar                               & 8.5            & 3.4 $\times 10^{-10}$                  & 2.9 $\times 10^{36}$\\ 
IGR J18410--0535 & SFXT                                        & 6.4            & 2.4 $\times 10^{-10}$                  & 1.2 $\times 10^{36}$ \\
Gamma Cas       & Be-HMXB pulsar                               & 0.17           & 1.9 $\times 10^{-10}$                  & 6.6 $\times 10^{32}$\\
4U 2206+543     & SFXT                                         & 3.4            & 1.9 $\times 10^{-10}$                  & 2.6 $\times 10^{35}$\\ 
Sco X--1        & Z type LMXB                                  &  2.8           & 2.9 $\times 10^{-7}$                   & 2.7 $\times 10^{38}$ \\
GRS 1915+105    & LMXB (BH)                                    &  12.5          & 2.4 $\times 10^{-8}$                   & 4.5 $\times 10^{38}$ \\
GX 5--1         & Z type LMXB                                  &  7.2           & 2.4 $\times 10^{-8}$                   & 1.5 $\times 10^{38}$ \\
GX 349+2        & Z type LMXB                                  &  8.5           & 1.8 $\times 10^{-8}$                   & 1.6 $\times 10^{38}$ \\
GX 17+2         & Z type LMXB                                  &  9.8           & 1.5 $\times 10^{-8}$                   & 1.7 $\times 10^{38}$ \\
GX 9+1          & Z type LMXB                                  &  7.2           & 1.4 $\times 10^{-8}$                   & 8.7 $\times 10^{37}$ \\
GX 340+0        & Z type LMXB                                  &  8.5           & 1.1 $\times 10^{-8}$                   & 9.5 $\times 10^{37}$ \\
Cyg X--2        & Z type LMXB                                  &  7.2           & 1.0 $\times 10^{-8}$                   & 6.2 $\times 10^{37}$ \\
GX 3+1          & Z type LMXB                                  &  5.0           & 9.1 $\times 10^{-9}$                   & 2.7 $\times 10^{37}$\\
GX 13+1         & atoll type LMXB                              &  7.0           & 7.3 $\times 10^{-9}$                   & 4.3 $\times 10^{37}$ \\
4U 1705--440    & LMXB                                         &  7.4           & 5.6 $\times 10^{-9}$                   & 3.7 $\times 10^{37}$ \\
GX 9+9          & Z type LMXB                                  &  7.0           & 4.7 $\times 10^{-9}$                   & 2.8 $\times 10^{37}$ \\
1A 1742--294    & LMXB                                         &  8.5           & 4.3 $\times 10^{-9}$                   & 3.7 $\times 10^{37}$ \\
Ser X--1        & atoll type LMXB                              &  8.4           & 4.2 $\times 10^{-9}$                   & 3.5 $\times 10^{37}$ \\
H 1735--444     & Z type LMXB                                  &  9.2           & 3.8 $\times 10^{-9}$                   & 3.8 $\times 10^{37}$ \\
4U 1630--472    & LMXB                                         &  4.0           & 2.0 $\times 10^{-9}$                   & 3.8 $\times 10^{36}$ \\
4U 1608--52     & atoll type LMXB                              &  4.0           & 1.4 $\times 10^{-9}$                   & 2.7 $\times 10^{36}$ \\
GS 1826--238    & atoll type LMXB                              &  6.7           & 1.2 $\times 10^{-9}$                   & 6.4 $\times 10^{36}$ \\
H 0614+091      & LMXB                                         &  3.0           & 1.1 $\times 10^{-9}$                   & 1.2 $\times 10^{36}$ \\
Her X--1        & LMXB pulsar                                  &  6.6           & 9.6 $\times 10^{-10}$                  & 5.0 $\times 10^{36}$ \\
SWIFT J1753.5--0127     & LMXB                                 &  6.0           & 9.3 $\times 10^{-10}$                  & 4.0 $\times 10^{36}$ \\
4U 1822--371    & LMXB                                         &  2.5           & 9.2 $\times 10^{-10}$                  & 6.9 $\times 10^{35}$ \\
4U 1746--37     & LMXB                                         &  16.0          & 9.1 $\times 10^{-10}$                  & 2.8 $\times 10^{37}$ \\
Aql X--1        & atoll type LMXB                              &  5.0           & 8.8 $\times 10^{-10}$                  & 2.6 $\times 10^{36}$ \\
Cir X--1        & atoll type LMXB                              &  10.9          & 8.8 $\times 10^{-10}$                  & 1.2 $\times 10^{37}$ \\
4U 1626--67     & LMXB                                         &  8.0           & 7.8 $\times 10^{-10}$                  & 5.9 $\times 10^{36}$ \\
GX 1+4          & LMXB pulsar                                  & 10.0           & 7.3 $\times 10^{-10}$                  & 8.8 $\times 10^{36}$ \\
GX 339--4       & LMXB (BH)                                    &  8.0           & 7.2 $\times 10^{-10}$                  & 5.5 $\times 10^{36}$ \\
4U 1543--624    & LMXB                                         &  10.0          & 7.2 $\times 10^{-10}$                  & 8.6 $\times 10^{36}$ \\
4U 1254--690    & LMXB                                         &  15.5          & 7.1 $\times 10^{-10}$                  & 2.0 $\times 10^{37}$ \\
Terzan 2        & LMXB                                         &  5.0           & 6.3 $\times 10^{-10}$                  & 1.9 $\times 10^{36}$ \\
HETE J1900.1--2455    &  LMXB                                  &  3.6           & 6.1 $\times 10^{-10}$                  & 9.5 $\times 10^{35}$ \\
H 1822--000     & LMXB                                         &  4.0           & 5.9 $\times 10^{-10}$                  & 1.1 $\times 10^{36}$ \\
4U 1957+115     & LMXB (BHC)                                   &  7.0           & 5.6 $\times 10^{-10}$                  & 3.8 $\times 10^{36}$ \\
SLX 1735--269   & LMXB                                         &  5.6           & 4.5 $\times 10^{-10}$                  & 1.7 $\times 10^{36}$  \\
4U 1954+319     & LMXB                                         &  1.7           & 3.9 $\times 10^{-10}$                  & 1.3 $\times 10^{35}$ \\
1H 1556--605    & LMXB                                         &  10.0          & 3.2 $\times 10^{-10}$                  & 3.8 $\times 10^{36}$ \\
4U 1323--619    & LMXB                                         &  11.0          & 3.0 $\times 10^{-10}$                  & 4.3 $\times 10^{36}$ \\
1H 0918--548    & LMXB                                         &  4.0           & 2.4 $\times 10^{-10}$                  & 4.6 $\times 10^{35}$ \\
MAXI J0556--332 & LMXB                                         &  8.5           & 1.8 $\times 10^{-10}$                  & 1.6 $\times 10^{36}$ \\
\hline
\end{tabular}
 \end{table*}
 
\subsection{Long term average spectra of X-ray binaries}

\begin{figure*}
\centering
\includegraphics[scale=0.25,angle=-90]{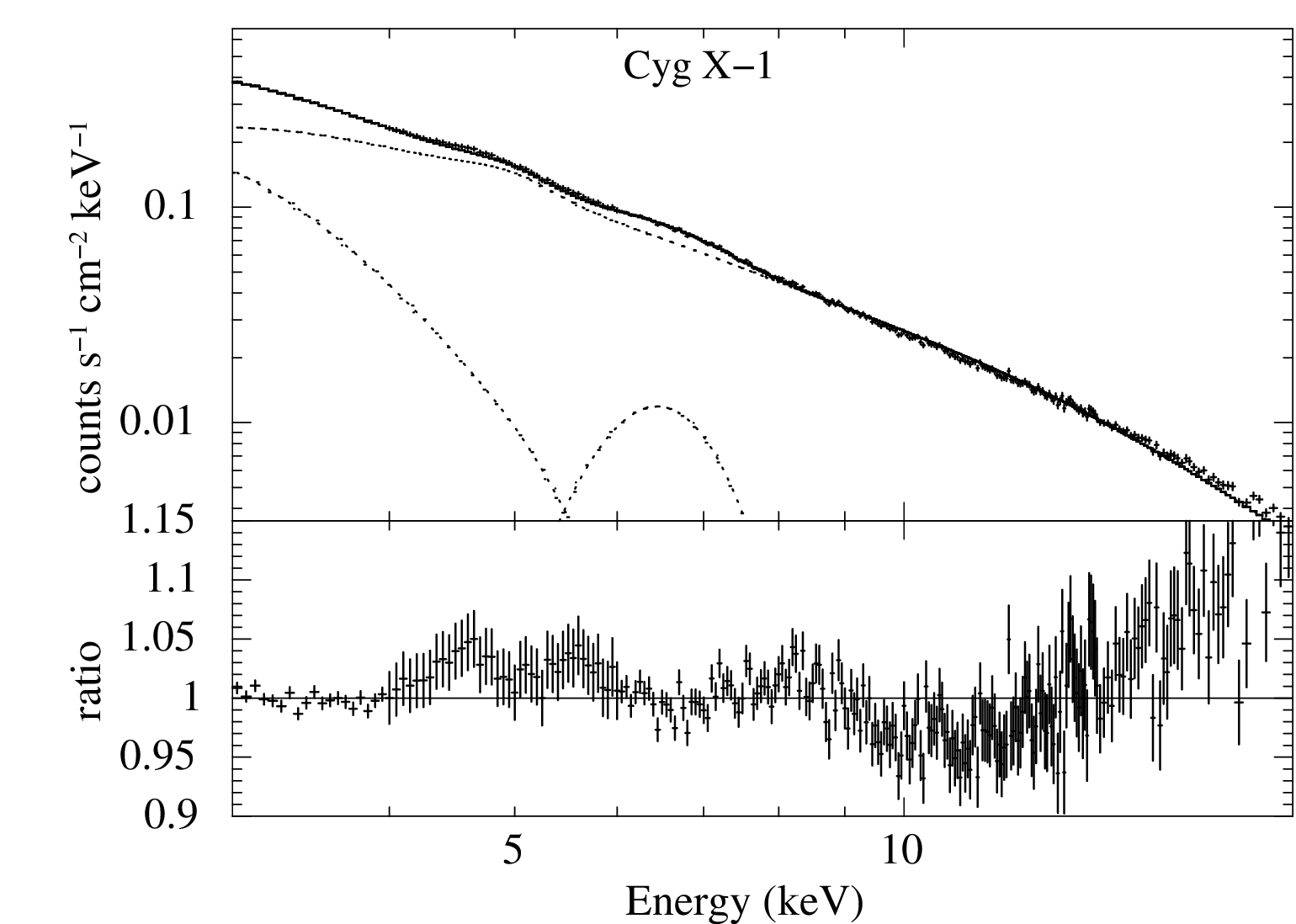}
\includegraphics[scale=0.25,angle=-90]{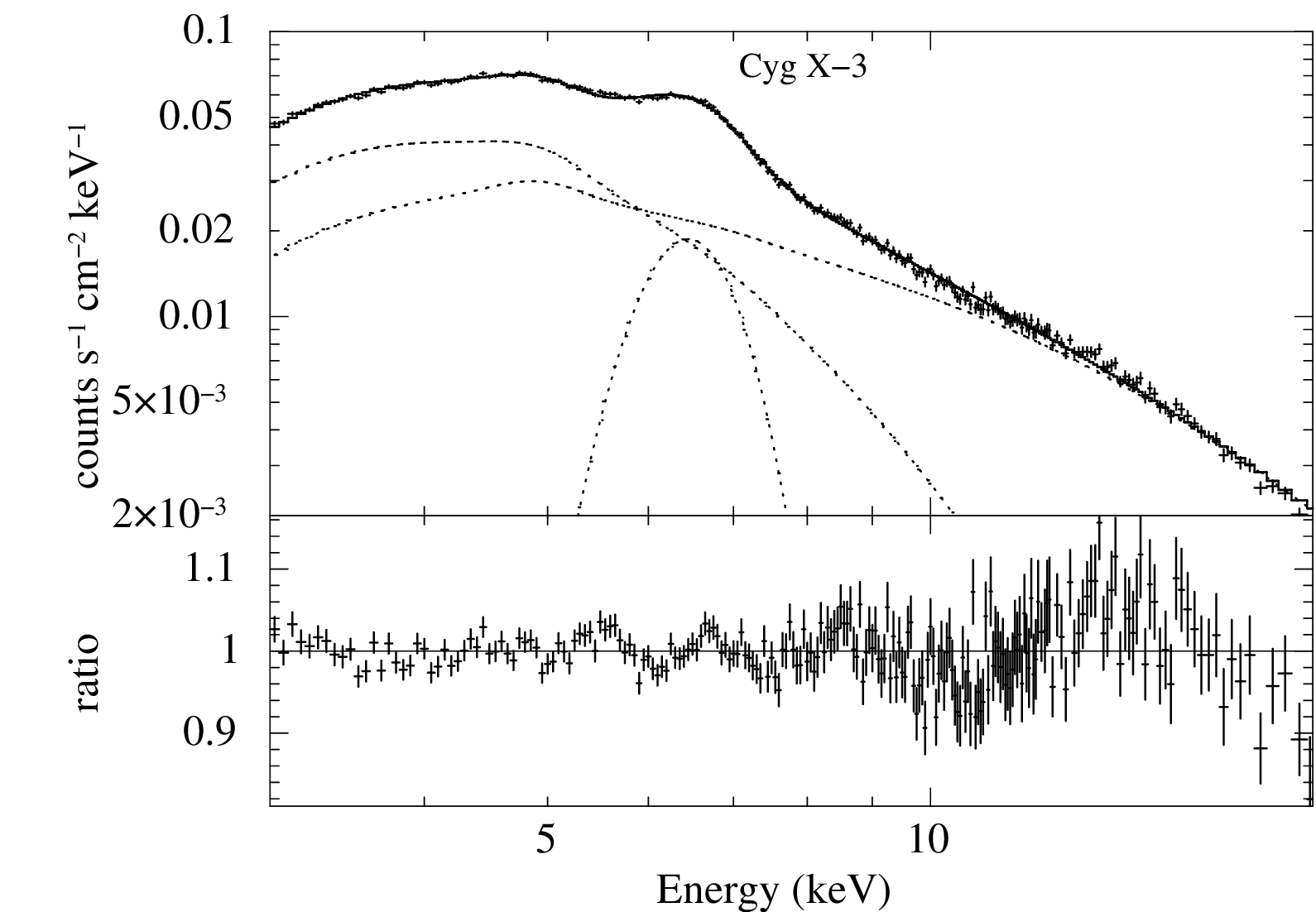}
\includegraphics[scale=0.25,angle=-90]{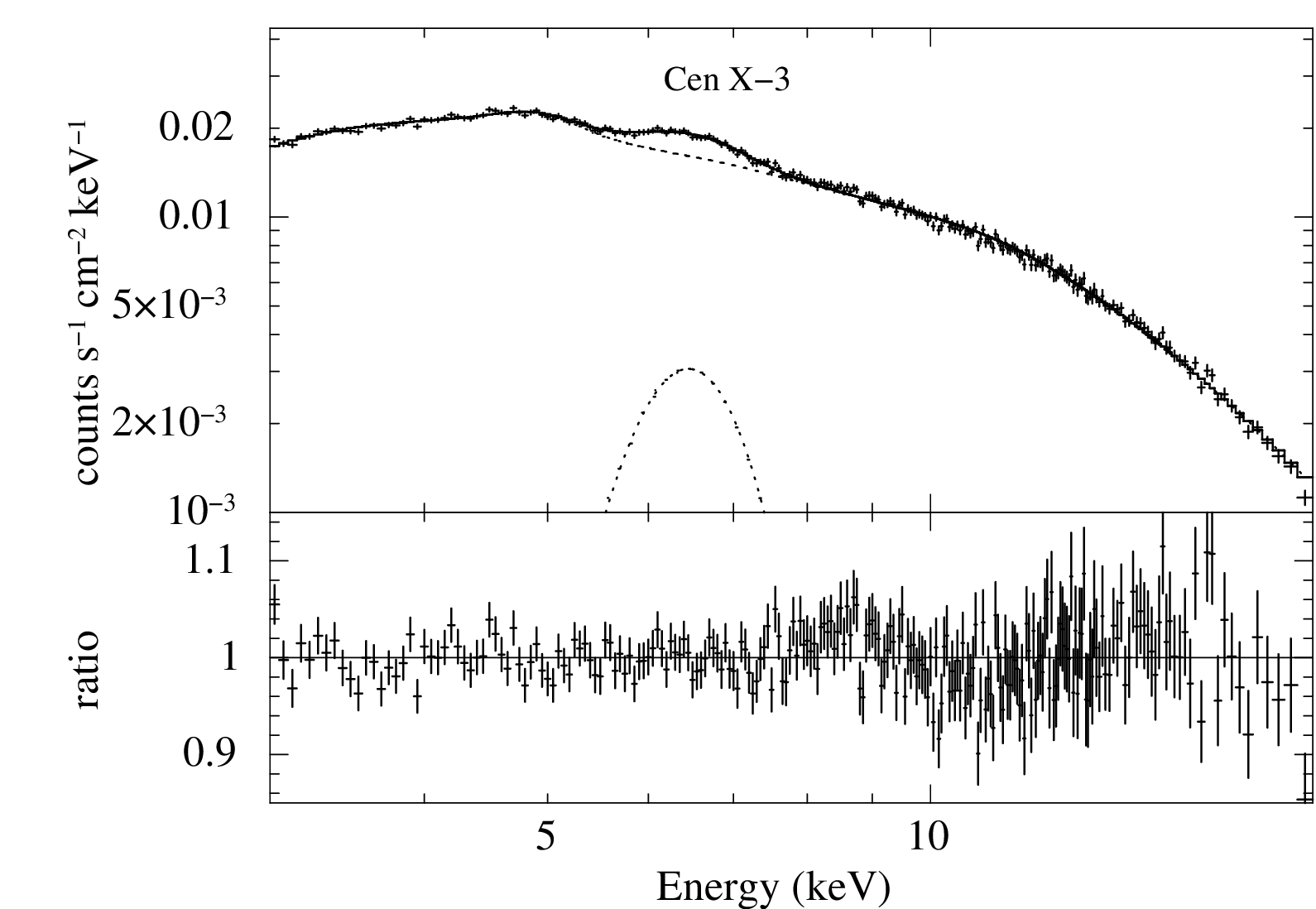}
\includegraphics[scale=0.25,angle=-90]{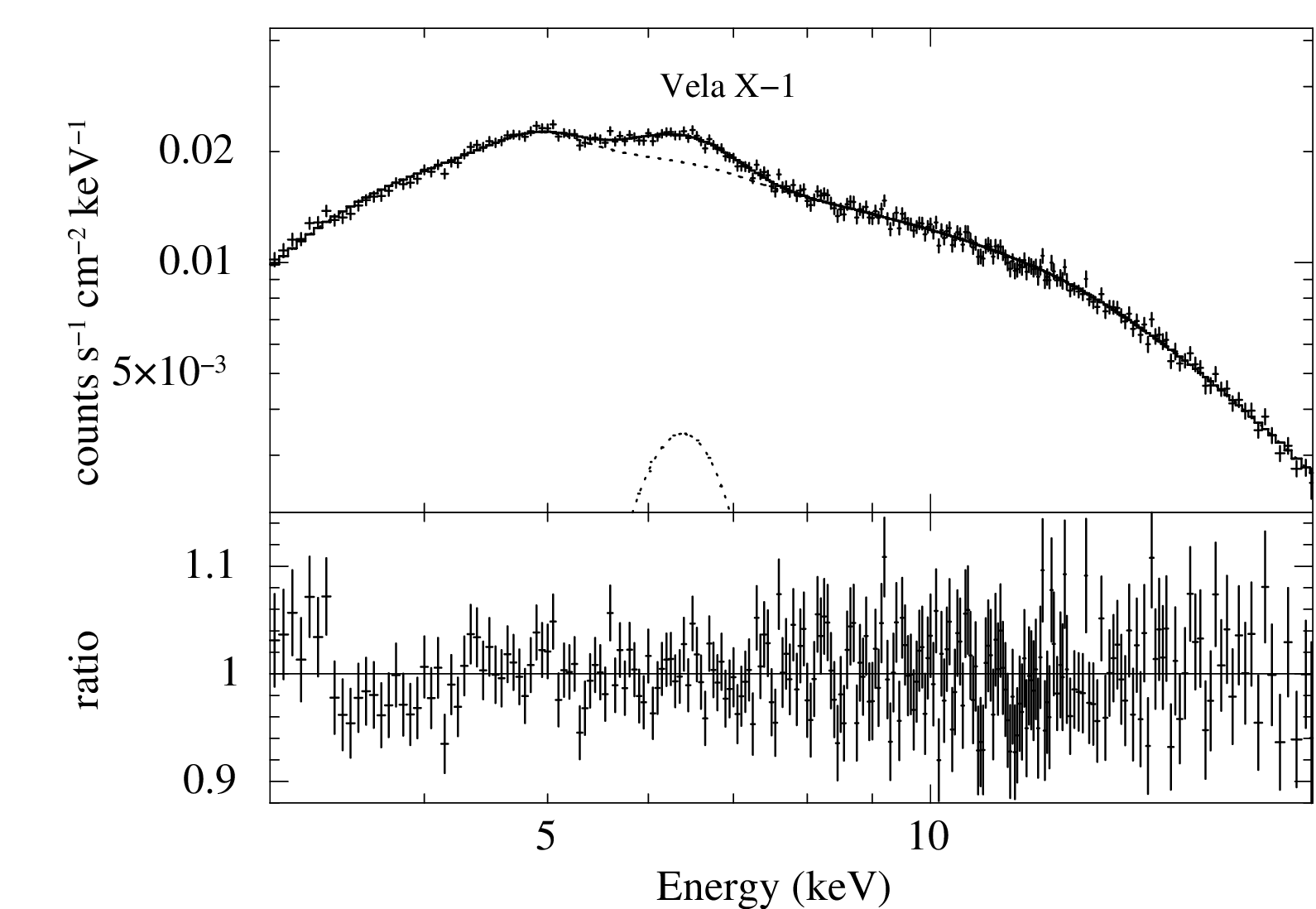}
\includegraphics[scale=0.25,angle=-90]{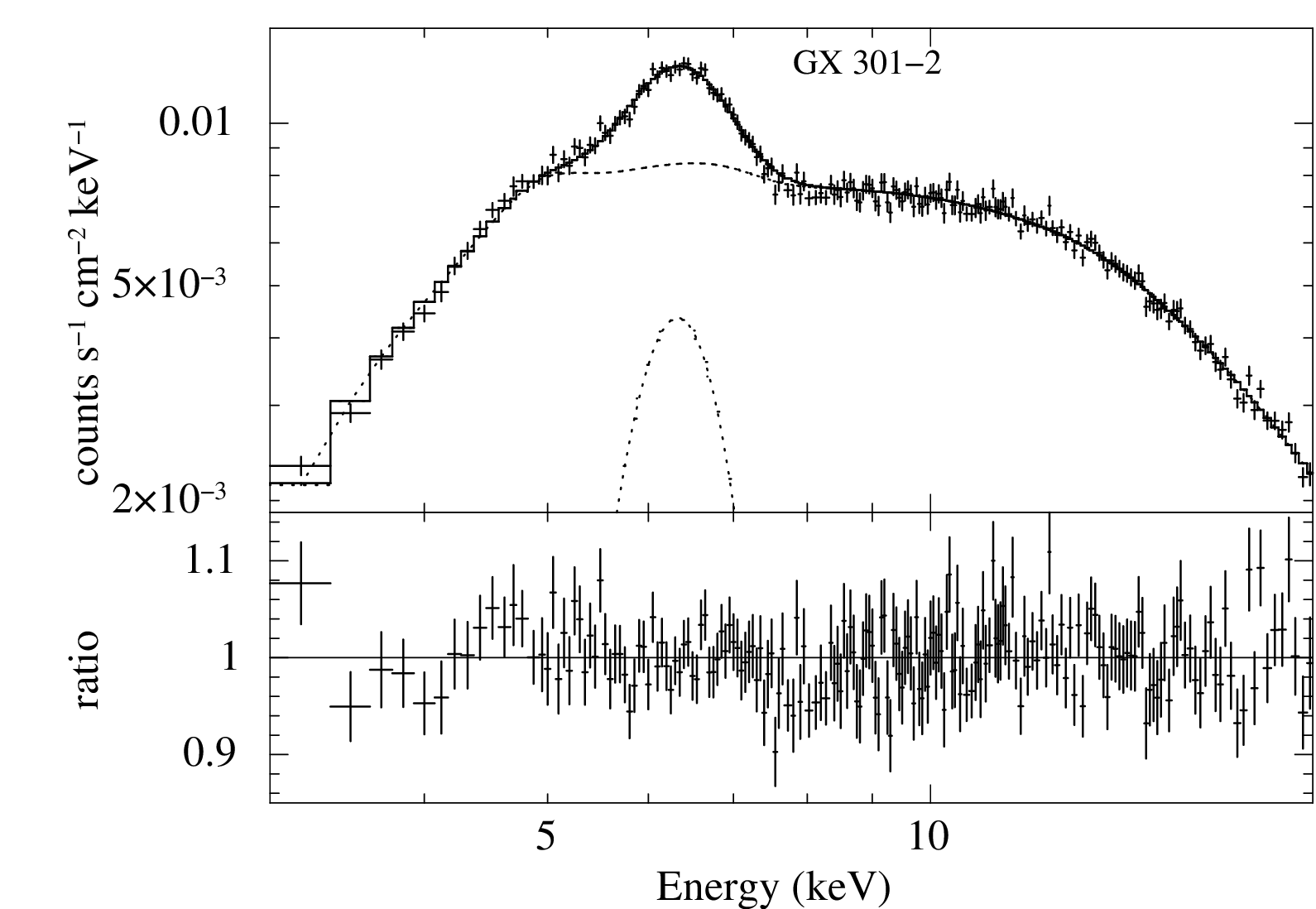}
\caption{Long term average spectra of 5 galactic HMXBs having the highest X-ray average flux, along with the best fit model and ratio of data and the best fit model.}
\label{hmxb_spectra}
\end{figure*}

\begin{figure*}
\centering
\includegraphics[scale=0.25,angle=-90]{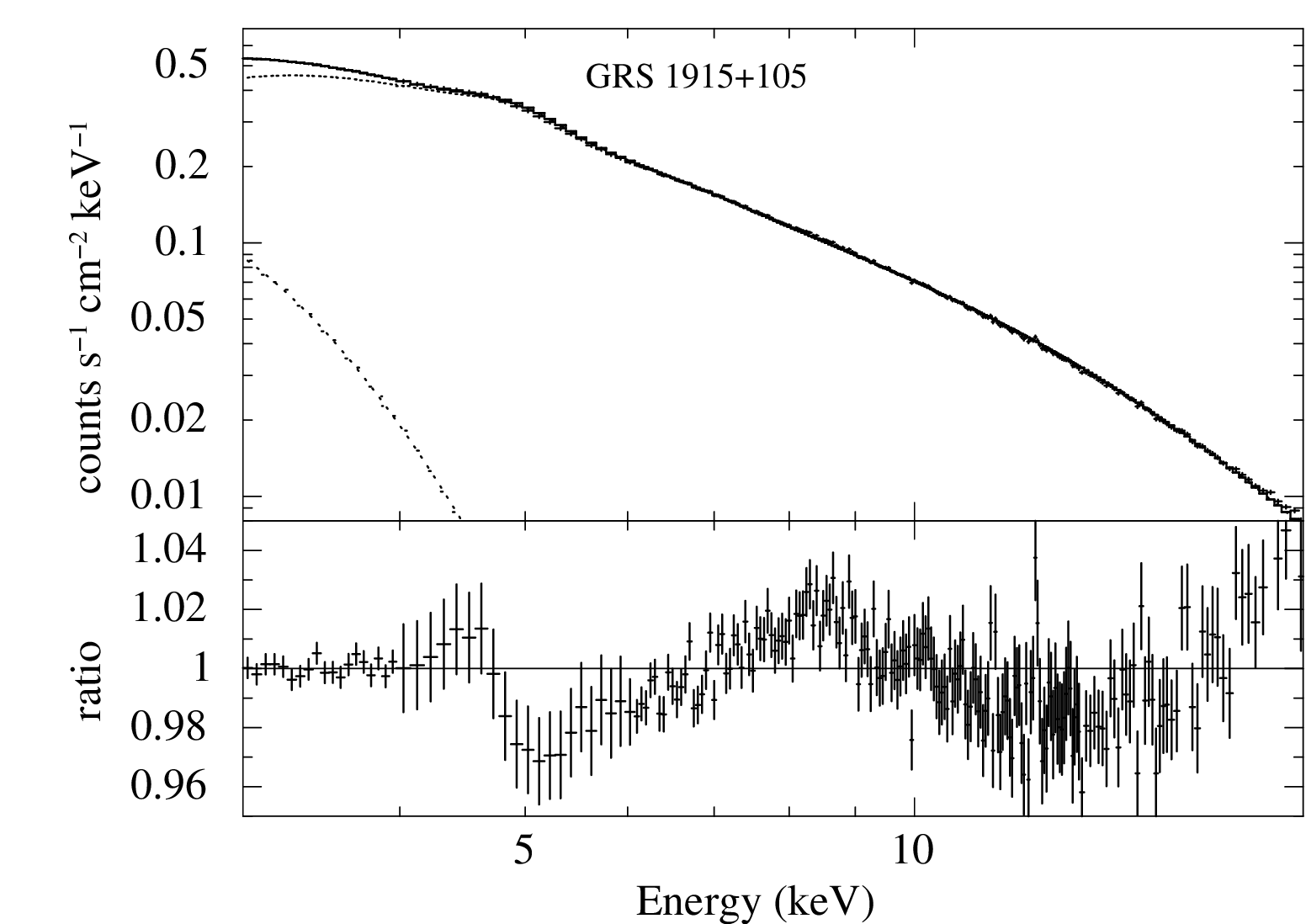}
\includegraphics[scale=0.25,angle=-90]{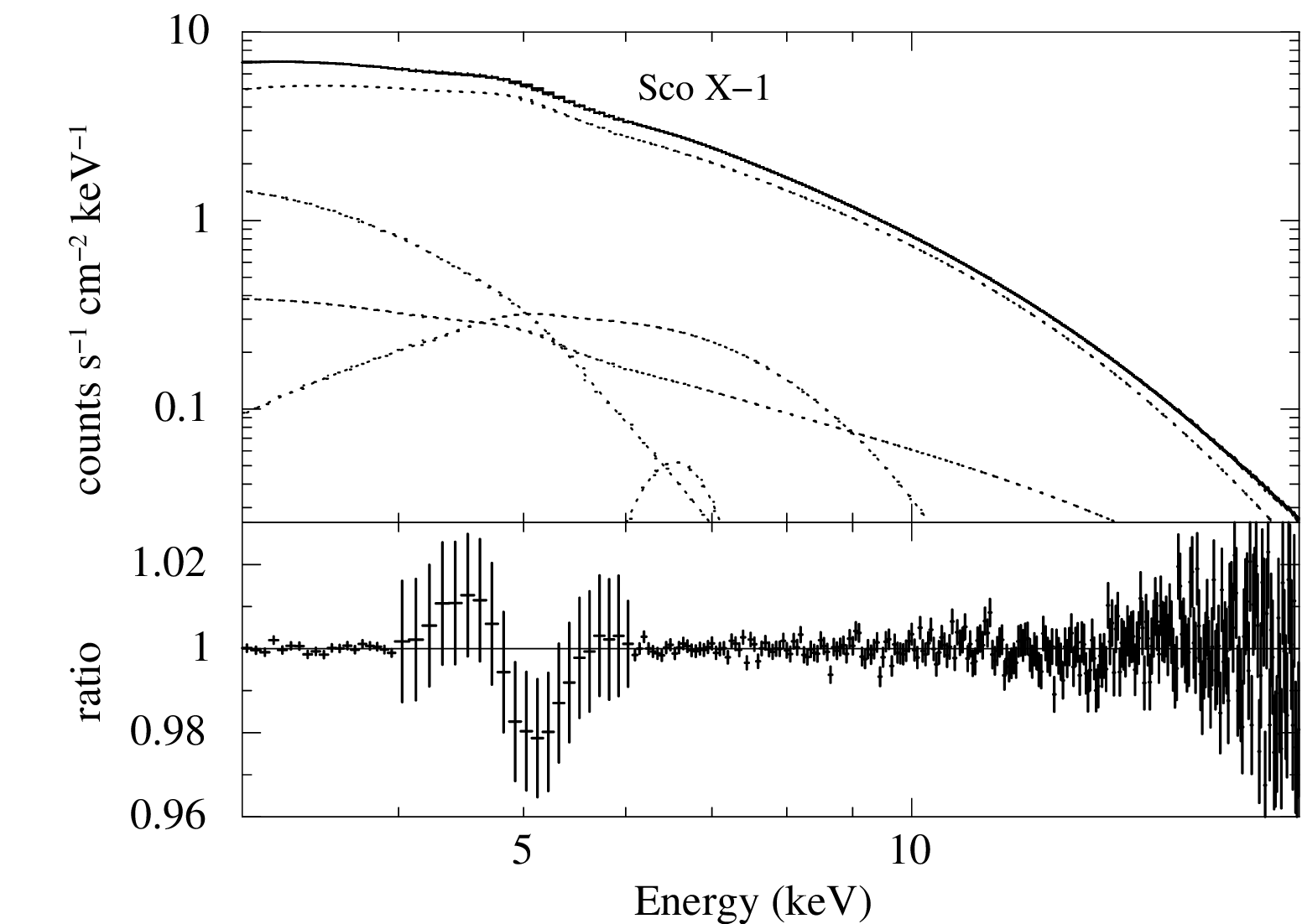}
\includegraphics[scale=0.25,angle=-90]{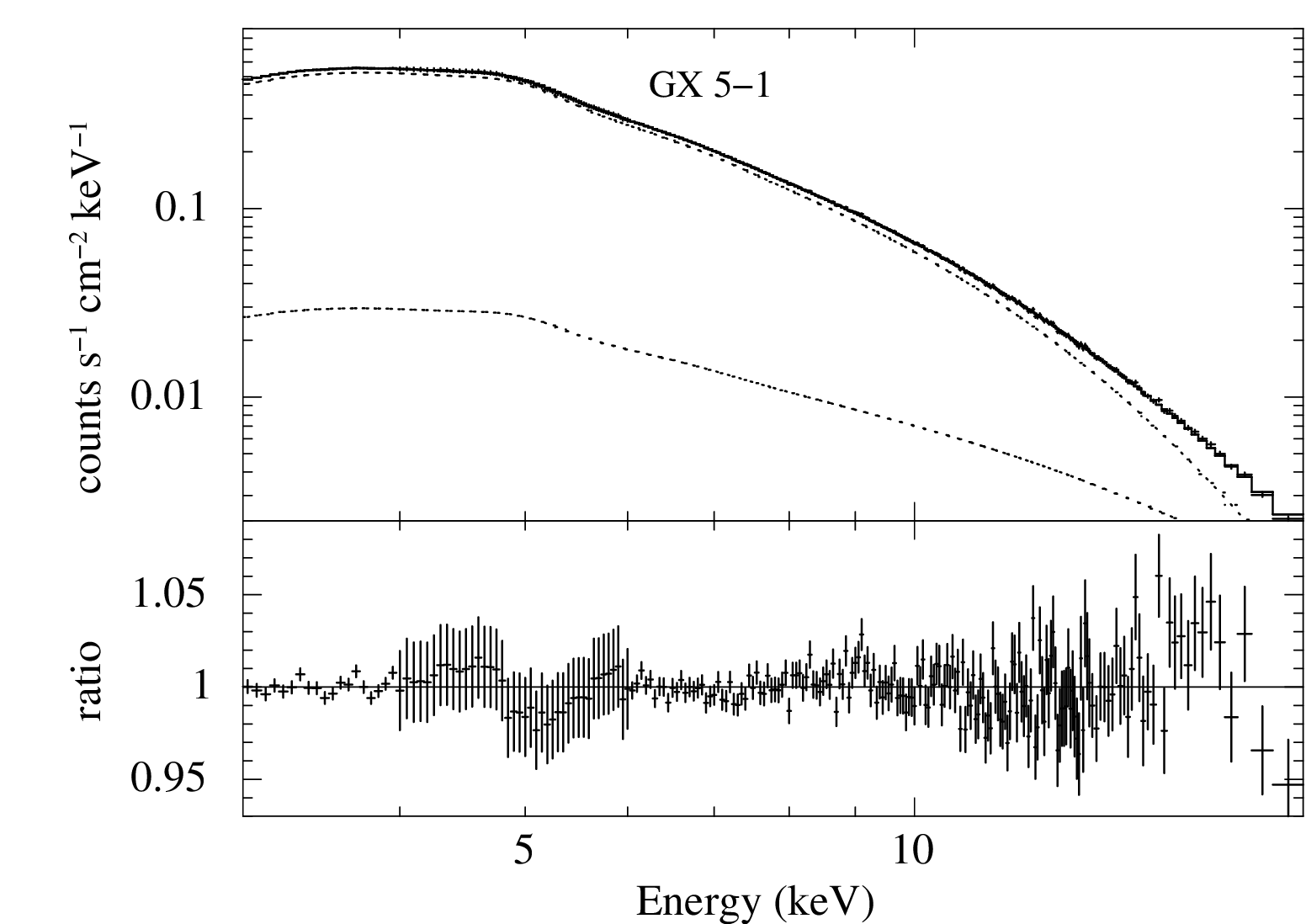}
\includegraphics[scale=0.25,angle=-90]{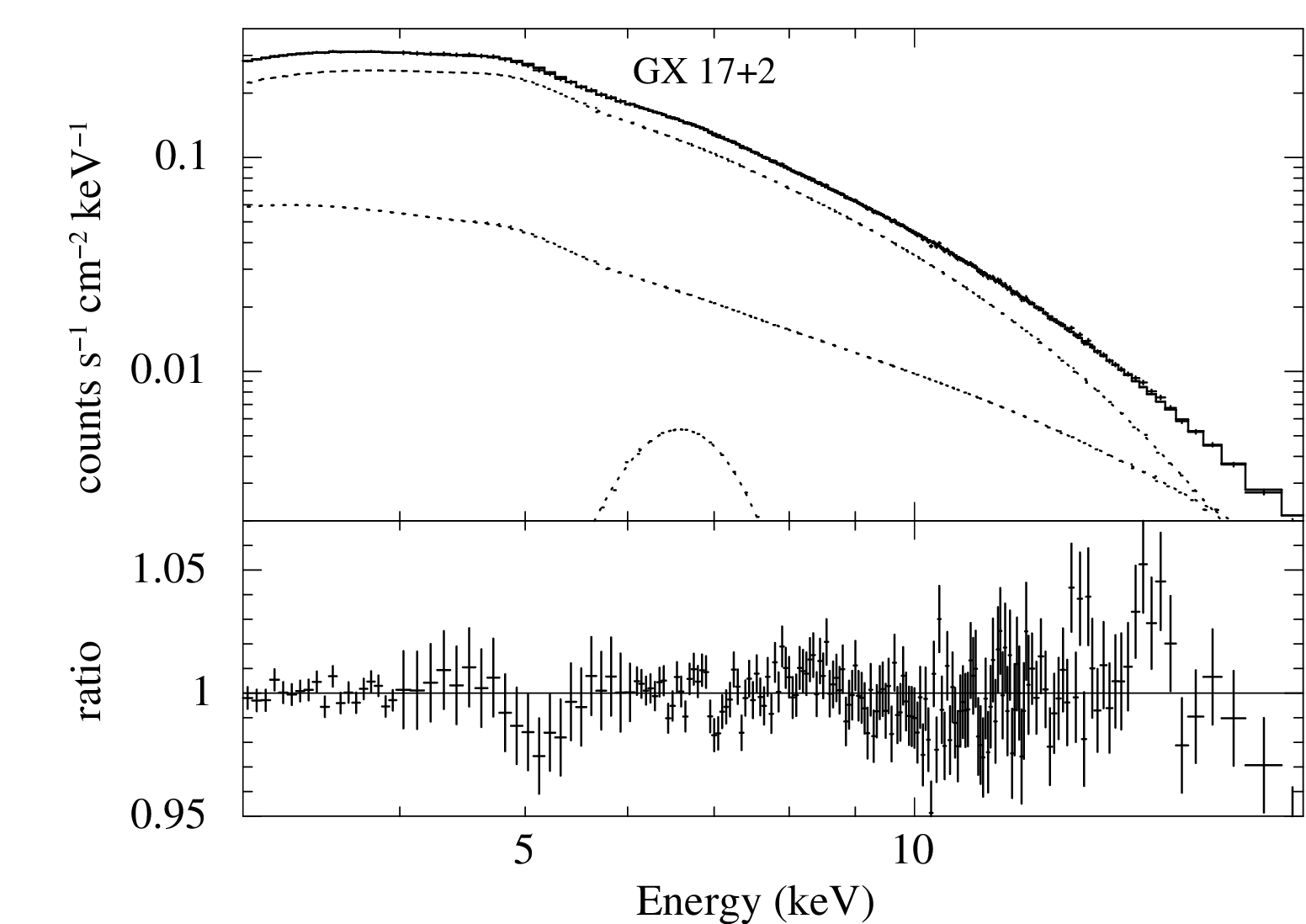}
\includegraphics[scale=0.25,angle=-90]{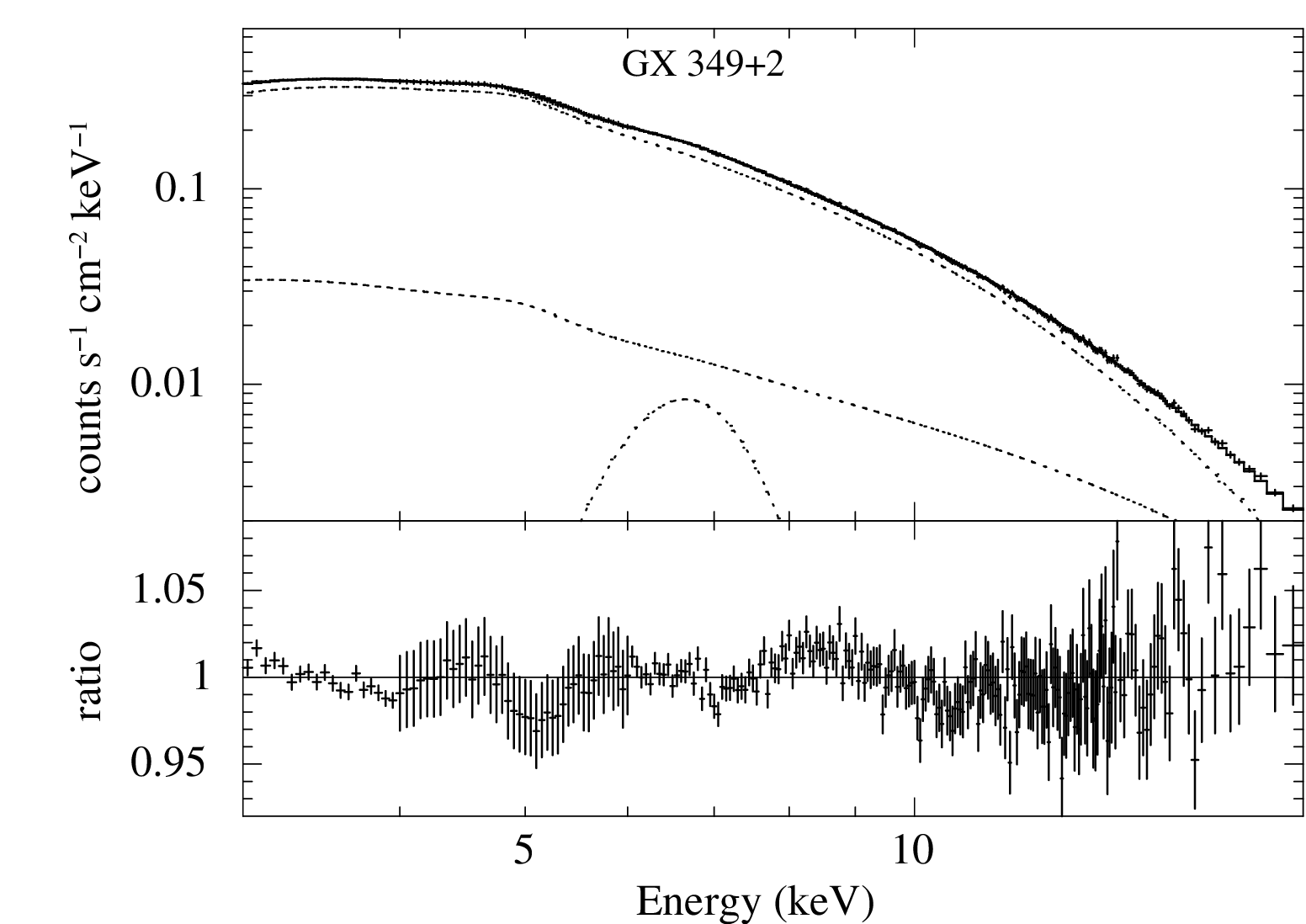}
\caption{Long term averaged spectra of 5 galactic LMXBs having the highest X-ray average flux, along with the best fit model and ratio of data and the best fit model.}
\label{lmxb_spectra}
\end{figure*}

The long term average spectra of X-ray binaries are fitted with most generally used phenomenological models in the energy range 3--20 keV, to estimate their average X-ray flux. For the cases where the photo-electric absorption cannot be constrained from the spectra, the value of $N_H$ is fixed to the column density of interstellar Galactic \HI along the line of sight (DL-nH) \footnote{nH tool in HEASARC: http://heasarc.gsfc.nasa.gov/cgi-bin/Tools/w3nh/w3nh.pl}. The X-ray spectral analysis were performed using {\small XSPEC} v:12.9 \citep{arnaud1996}. The various X-ray spectral models mentioned below are specified in {\small XSPEC} \footnote{https://heasarc.gsfc.nasa.gov/xanadu/xspec/manual/Models.html}.
\par
Majority of the galactic HMXBs are accretion powered X-ray pulsars. Observations of Milky Way, LMC and SMC show that the population of HMXBs is dominated by the Be-HMXBs \citep{negueruela2002,reig2011,haberl2016}. Be--HMXBs are transients; they are mostly in quiescence with very low X-ray luminosity and occasionally go into outbursts. Hence their average X-ray spectra is related to their duty cycle of outbursts in $\sim$5 years of MAXI operation. Except for the Be--HMXBs listed in Table 1, majority of the Be--HMXBs have low average X-ray flux. The other HMXBs in Table 1, are the bright and persistent HMXBs. The X-ray spectra of these pulsars are generally modelled by a power-law with a high energy cut-off, modified by photo-electric absorption by the column density of absorbing matter along the line of sight \citep{white1983,maitra2013,islam2014,islam2015}. We have fitted the long term average spectra of these pulsars with the above spectral model. The X-ray spectra of black hole binaries consists of a thermal component, modelled by a multi-colour disk blackbody `diskbb' model \citep{mitsuda1984} or a blackbody model `bbody', and a non-thermal component, modelled by a thermal Comptonisation model `nthcomp' \citep{zdziarski1996}. The spectra of LMXBs are modelled with a multi-colour disk blackbody `diskbb' and a power-law model. A Fe K$\alpha$ line is found to be present in the spectra of most of the X-ray binaries, and it is modelled with a gaussian emission line. 
\par
The long term averaged spectra of the 5 galactic HMXBs with the highest X-ray average flux: Cyg X--1, Cyg X--3, Cen X--3, Vela X--1 and GX 301--2, along with their best fit models and ratio of the data to the model is shown in Figure ~\ref{hmxb_spectra}. The long term averaged spectra of the 5 galactic LMXBs with the highest X-ray average flux: Sco X--1, GRS 1915+105, GX 5--1, GX 17+2 and GX 349+2, along with their best fit models and ratio of data to the model are shown in Figure ~\ref{lmxb_spectra}. Detailed studies of the orbital dependence of the spectral parameters have been carried out with {\it MAXI}--GSC for GX 301--2 \citep{islam2014}, Vela X--1 \citep{doroshenkov2013} and 4U 1538--52 \citep{roca2015}. 

\subsection{Construction of composite X-ray binary spectra}

\begin{figure*}
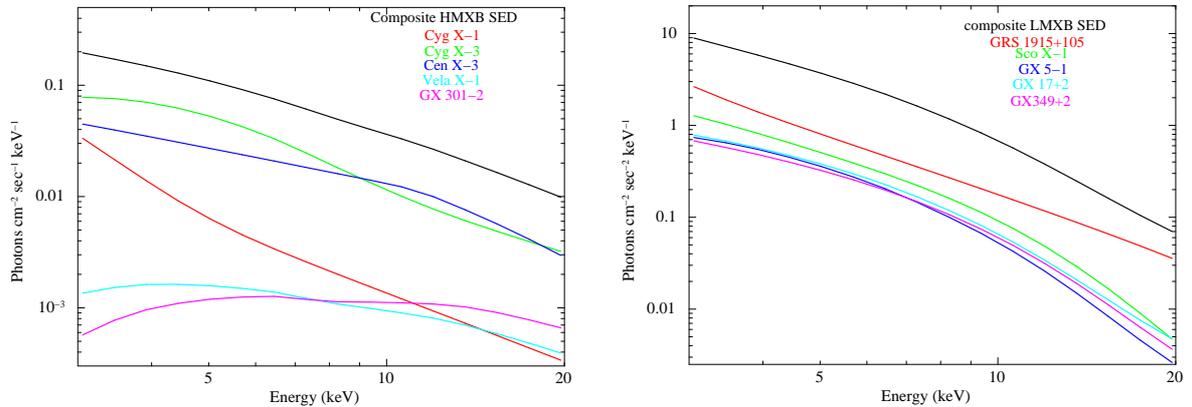

\centering
\includegraphics[angle=-90,scale=0.3]{overlay_hmxb.ps}
\includegraphics[angle=-90,scale=0.3]{overlay_lmxb.ps}
\caption{Composite HMXB and LMXB spectra constructed from the long term averaged spectra of X-ray binaries with {\it MAXI}--GSC data, taking into account the circumstellar absorption. Overlaid on the composite spectra are the spectra of 5 HMXBs and LMXBs having the highest X-ray average fluxes, normalised to the Galactic centre distance, showing their contribution to the composite X-ray binary spectra.}
\label{galaxy_sed}
\end{figure*}

We use the long term average spectra of galactic X-ray binaries (described in previous subsection) to construct the composite X-ray binary spectra of HMXBs and LMXBs separately, after accounting for circumstellar absorption. The Fe fluorescence lines are produced by the reprocessing of the radiation by the surrounding matter and is more related to the property of the reprocessing matter than the source spectrum. To construct the composite spectra from X-ray binaries, we have removed these Fe fluorescence lines from the spectra. The total photo-electric absorption consists of column density of interstellar Galactic \HI~along the line of sight, along with the circumstellar matter around the X-ray binary. We have removed the effect of the interstellar Galactic \HI~from the total photo-electric absorption. The resultant spectra are scaled to the Galactic centre distance of 8.5 kpc and then added up to construct composite X-ray binary spectra. One of our goals in this paper is to determine the implication of the X-ray binary spectra for the high redshift intergalactic medium and for this purpose the interstellar absorption along the Galactic disk is not relevant. Figure ~\ref{galaxy_sed} shows the composite X-ray binary spectra constructed separately for galactic HMXB and LMXB population. Overlaid on the composite HMXB SED (left panel) are the spectra of 5 HMXBs with the highest X-ray average flux: Cyg X--1, Cyg X--3, Cen X--3, Vela X--1 and GX 301--2, normalised to the Galactic centre distance. Also overlaid on the composite LMXB SED (right panel) are the spectra of 5 LMXBs with the highest X-ray average flux: GRS 1915+105, Sco X--1, GX 5--1, GX 17+2 and GX 349+2, normalised to the Galactic centre distance. The composite spectra of both HMXBs and LMXBs can be approximated with three piece-wise power-law functions, with steeper photon indices at higher energies.
\par
The composite HMXB SED is described empirically by the following function in the energy range of 3--20 keV.
\begin{eqnarray}
 \label{hmxb_empirical}
N(E) &=& 0.68 \times E^{-1.11}~\mathrm{for}~3 < E \leq 5~\rm keV \nonumber \\
     &=& 1.44 \times E^{-1.59}~\mathrm{for}~5 < E \leq 10~\rm keV \nonumber \\
     &=& 3.01 \times E^{-1.91}~\mathrm{for}~10 < E < 20~\rm keV.
\end{eqnarray}

The composite LMXB SED is described empirically by the following function in the energy range of 3--20 keV.
\begin{eqnarray}
 \label{lmxb_empirical}
N(E) &=& 68.3 \times E^{-1.81}~\mathrm{for}~3 < E \leq 7~\rm keV \nonumber \\
     &=& 439.9 \times E^{-2.81}~\mathrm{for}~7 < E \leq 12~\rm keV \nonumber \\
     &=& 1943.5 \times E^{-3.44}~\mathrm{for}~12 < E < 20~\rm keV.
\end{eqnarray}

%%%%%%%%%%%%%%%%%%%%%%%%%%%%%%%
\section{Implications in Epoch of Reionization}
\label{sec:eor}
Now, we will use the estimated composite HMXB spectra to investigate the impact of HMXBs on the redshifted 21-cm signal from the Cosmic Dawn and EoR. We have used a 1D radiative transfer (RT) simulation {\sc grizzly} \citep{ghara15a, 2018MNRAS.476.1741G} to generate the 21-cm signal maps using the outputs of an $N$-body simulation. The following sections describe the simulations used in this study and the outcomes from this investigation. 

%%%%%%%%%%%%%%%
\subsection{Simulations}
\label{sec:sim}
We briefly present the $N$-body simulation, source model and 1D RT simulation used to generate the 21-cm signal maps from the Cosmic Dawn and EoR in this Section and describe the results in Section \ref{sec:res}.

%%%%%%%%%%%%%%%%
\subsubsection{$N$-body simulation}
\label{sec:nbodysim}
The density and velocity fields used in this paper are generated using the publicly available $N$-body simulation code {\sc cubep}$^3${\sc m}\footnote{\tt http://wiki.cita.utoronto.ca/mediawiki/index.php/CubePM} \citep{Harnois12}. 
The details of the dark matter only simulation are: (i) the number of dark-matter particles is $1728^3$, (ii) number of fine grids used in the simulation is $3456^3$, while the density and velocity fields are generated in a grid 8 times coarser than the fine grids, (iii) simulation box size is 200 $h^{-1}$ comoving megaparsec (cMpc), (iv) mass of the dark matter particles is $2 \times 10^8\,  \MSUN$. We generate the density and velocity fields within the redshift range $25 \geq z \geq 6$ in snapshots such that the time gap between two nearby snapshots is 10 megayear (Myr). The baryonic density and velocity fields are assumed to follow the dark matter fields. The dark matter halos in the simulation box are identified using the spherical overdensity method. The mass resolution of the simulation limits us properly resolving the halos with mass $\lesssim 8 \times 10^9\, \MSUN$. The minimum mass of the halos considered in this study is $\sim 8 \times 10^9\, \MSUN$ which corresponds to a virial temperature of $2.7\times 10^5$ K at redshift 6. Throughout the paper, we have used the cosmological parameters $\Omegam=0.32$, $\Omega_\Lambda=0.68$, $\OmegaB=0.049$, $h=0.67$, $n_{\rm s}=0.96$, and $\sigma_8=0.83$ \citep{Planck2013}.

%%%%%%%%%%%%%%%%%
\subsubsection{Radiating sources}
\label{p5rs}

\begin{figure*}
\begin{center}
\includegraphics[scale=0.75]{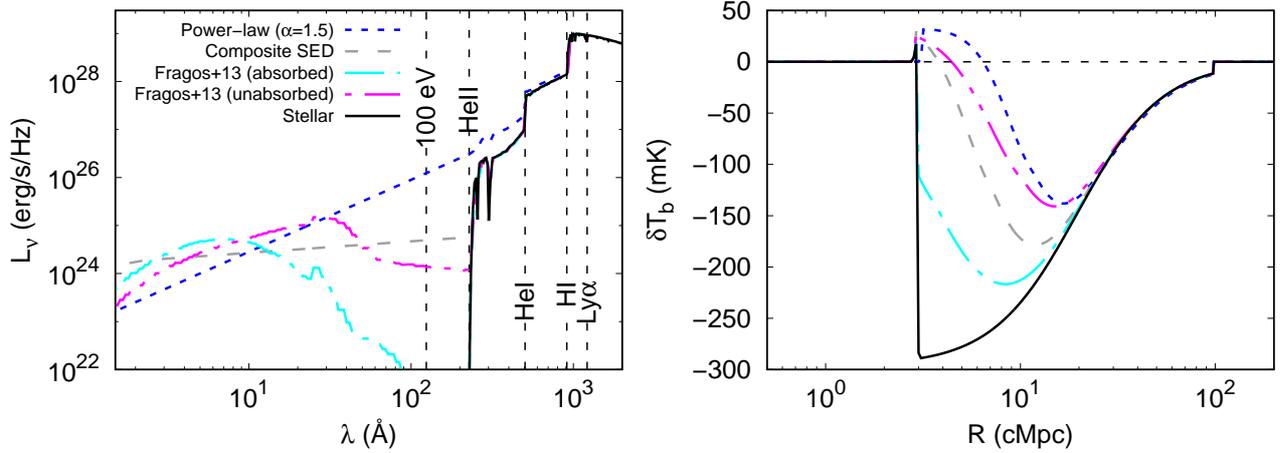}
    \caption{{{\bf Left panel:} The short-dashed curve represents the power-law spectrum of the X-ray part of the mini-QSO type sources, while the long-dashed curve shows the composite spectrum of the HMXBs used in this paper. The solid curve represents the stellar contribution from the galaxy     for a stellar mass of $10^8 ~\MSUN$.  For comparison, we have also plotted the intrinsic HMXB spectrum (magenta double-dot dashed curve) and the absorbed HMXB spectrum (cyan dot-dashed curve) from \citet{fragos2013}. The absorbed spectrum includes absorption of the soft X-ray photons emitted from the HMXB within the host galaxy. Note that we have fixed the parameter $f_X$ (the ratio of the luminosities in the UV and X-ray bands) as 0.05 for the source model with X-rays while we choose the spectral index of the mini-QSO SED to be 1.5.} The right to left vertical lines represents the energy correspond to $\lya$, the ionization energy of     \HI, \HeI, \HeII and 100 eV respectively. {\bf Right panel :} The brightness temperature profiles as a function of the radial distance from the     centre of the sources presented in the left panel of the figure. The stellar mass of the sources is assumed to be $10^8 ~\MSUN$, while the profiles correspond to 20 Myr source age at redshift 15.  }
   \label{image_p5spectrum}
\end{center}
\end{figure*}

We assume that the dark matter halos contain the main sources of ionizing photons, which is assumed to be the stars in the galaxies. We assume that the stellar mass of a galaxy hosted in a dark matter halo of mass $M_{\rm halo}$ is $M_\star=f_\star \left(\frac{\OmegaB}{\Omegam}\right) M_{\rm halo}$. The parameter $f_\star$ represents the fraction of the baryon in the form of stars residing within a galaxy. Given the SED of the sources, the reionization 
history depends on $f_\star$. We choose the value of the parameter $f_\star$ such that the reionization history is consistent with the recent estimation of the Thomson scattering optical depth from the cosmic microwave background radiation (CMBR) observations.

The UV and near-infrared SED of the stellar contribution from the galaxies are generated using the publicly available stellar population synthesis code {\sc pegase2}\footnote{\tt http://www2.iap.fr/pegase/} \citep{Fioc97}. We assume that the galaxies follow a Salpeter IMF for the stars within the mass range 1 to 100 $\MSUN$. We assume the metallicity to be 0.1 $Z_\odot$ throughout the reionization period. In this model, the SED of the sources 
assumed to have same shape throughout the reionization history and only scale with the stellar mass of the sources. In addition to that, we assume a fraction (1-$f_{\rm esc}$) of the UV photons in the intrinsic spectrum, absorbed in the interstellar medium (ISM). $\lya$ photons considered in this study are contributed by the continuum SED and the recombination of the absorbed ionizing photons in the ISM.

In addition to the stellar contribution of the ionizing photons, there could be contributions from the mini-QSO and HMXBs-type sources. While the major part  of the ionizing photons comes from the stellar part of the SED, the X-ray photons from mini-QSO and HMXBs can partially ionize and heat up the neutral region. In this paper, we model the X-ray part from the mini-QSO as a power-law SED which can be written as, 
\begin{equation}
I_q(E) = A_q ~ E^{-\alpha}, 
\label{eq_Iq}
\end{equation}
where $\alpha$ is the spectral index. We define a parameter $f_X$ which is the ratio of the  X-ray and UV luminosities. The normalization constant $A_q$ in Equation. \ref{eq_Iq} can be determined in terms of this parameter $f_X$. In this study, we choose $f_X=0.05$ and $\alpha = 1.5$ and called the source model as ``power-law'' source model.

To model the X-ray binaries spectra, we used the composite HMXBs spectra given in the functional form in Equation \ref{hmxb_empirical}. This functional form of the composite HMXB SED is extrapolated to lower energies. The normalisation of these composite HMXBs SED can be fixed by the parameter $f_X$ used in this study. We fix $f_X = 0.05$  and denote this as ``composite'' spectrum. Note that the extrapolation of the spectrum from high energy side to the soft X-ray end is rather an ad-hoc approximation. In principle, the soft X-rays are more probable to be absorbed in the interstellar medium than the hard X-rays. In addition, the absorption of the soft X-ray photons increases with the increase in metallicity in the host galaxies. However, these high redshift galaxies are expected to be less metal enriched \citep[see e.g.,][]{2011MNRAS.414..847S, 2012ApJ...745...50W, das2017} during the Cosmic Dawn which justifies the extrapolation.  On the other hand, the X-ray emission from hot ionised gas having temperatures of 0.1--0.2 keV dominates the soft X-rays \citep{mineo2012, pacucci2014}. Since MAXI/GSC operates in the energy range of 3--20 keV, we extrapolated the power-law index of the composite HMXB spectra in 3--5 keV i.e an index of 1.1 to lower energies. This index is similar to that estimated by \citet{pacucci2014}, using Chandra observations of star-forming galaxies, predominantly consisting of HMXBs, in its operating range of 0.3--8.0 keV. However, most of the galactic HMXBs are accretion powered pulsars and the broad band spectrum, whenever measured with instrument covering low energy band, shows it to be a single power law. Only in a few cases, another soft component is detected, with temperature of about 0.1 keV that amounts to only about a percent of the flux. For example, using CCD spectra of 30 HMXBs covering energies less than 1 keV, \citet{pradhan2018} fitted all the spectra  with a single power-law, very few exceptions having small contribution from a soft component.
Therefore, extrapolating the long term data which is above 3 keV, to lower energies is justified.

The left panel of Figure \ref{image_p5spectrum} shows the two SEDs, namely the power-law and composite SED, for stellar mass $M_{\star}=10^8 ~\MSUN$ which corresponds to a dark matter halo with mass $\sim 3.3\times 10^{10}\MSUN$.  For comparison, we have also plotted the intrinsic (magenta double-dot dashed curve) and absorbed (cyan dot-dashed curve) HMXB spectrum from \citet{fragos2013} in the left panel of this Figure\footnote{Note that the absorbed and unabsorbed spectrum from \citet{fragos2013} have the same number of photons at wavelength $\lambda$ $\lesssim$ 6  $\mathrm{\AA}$ and only deviate at higher wavelength side. In our case, the absorbed and unabsorbed spectrum differ at low wavelength side as we have normalized both the spectrum so that the ratios of luminosities in the X-ray and UV bands for both the spectrum are  0.05. }. We have also chosen $f_X=0.05$ for these HMXB spectra.  While the intrinsic HMXB spectrum contains a significant amount of soft X-ray photons,  one can easily notice that the absorbed HMXB spectrum from \citet{fragos2013} contains fewer soft X-ray photons with energy $E\sim 0.1$ keV due to the absorption of the soft X-ray photons within the interstellar medium of the host galaxy.

%%%%%%%%%%%%%%%
\subsubsection{Signal Maps}
Here, we briefly explain the method used in this paper to generate the 21-cm maps using the outputs of the $N$-body simulation described in the previous section. The details of the method can be found in \citet{ghara15a}. The UV photons have small mean free path and thus, will be absorbed by the nearby neutral hydrogen in the IGM. This gives rise to a highly ionized \HII ~region just outside the sources.  On the other hand, the mean free path of the 
X-rays crucially depends on the energy of the X-rays. The mean free path can be given as \citep[e.g.][]{2012MNRAS.426.1349M}
\begin{equation}
{\lambda}_X \approx 34 ~\bar{x}_{\rm HI}^{-1} \left( \frac{E_X}{0.5 ~\rm keV}\right)^{2.6} \left( \frac{1+z}{15} \right)^{-2} \rm cMpc,
\label{equ_lamdaX}
\end{equation}
where $\bar{x}_{\rm HI}$ is the average neutral fraction of the IGM and $E_X$ is the energy of the X-ray photon . Thus, X-rays have longer mean free path and  are able to penetrate longer distances in the IGM. In presence of the X-rays, the ionization and heating morphologies in the neutral regions are expected to be different than the scenario where there are no X-rays in the SED. However, the amount of partial ionization and heating in the IGM 
mostly depends on the number of soft X-rays in the SED of the sources. Given a SED, the steps to generate the ionization and kinetic temperature maps are given below.

\begin{itemize}
\item We used a one-dimensional radiative transfer method to generate the ionization and kinetic temperature maps from the density fields and the halo list from the $N$-body simulation. First, we generate a large set of 1D profiles of $\XHII$ and $\TK$ around the sources for different stellar masses, age, redshift and IGM overdensity. These profiles are used to estimate the ionizing photons from the sources and create \HII ~bubbles. In case there are overlaps between the individual \HII ~bubbles, we increase the overlapping bubble size by redistributing the unused ionizing photons among the overlapping 
sources.

\item We use the previously generated 1D profiles of $\XHII$ to calculate the ionization fraction in the neutral medium in the IGM. Finally, we use a correlation of the kinetic temperature and the ionization fraction to generate the $\TK$ maps. \footnote{The mean-free-path of the X-ray photons with energy $\gtrsim$1 keV is much larger than our simulation box size. In principle, these high-energy X-ray photons should be recycled into the simulation box until those are absorbed by the gas after sufficient redshifting. However, the recycling of such high-energy X-ray photons is computation very expensive and thus, often ignored in earlier studies such as \citet{2018arXiv180803287R}. Eventually, these unabsorbed photons will produce a uniform background. Note that we have not accounted for such X-ray background or recycle such unabsorbed photons into the simulation box. Our composite spectrum contains a significant amount of soft X-rays and thus, the heating is dominated by the soft X-rays.}

\item We have not performed any radiative transfer for the $\lya$ photons in this study and simply assumed that the $\lya$ photon flux reduce as $1/R^2$, where $R$ is the radial distance from the source. Using the $\lya$ flux, we calculate the $\lya$ coupling coefficients which later are used to generate the spin temperature ($\TS$) maps using the $\TK$ maps. All these maps are finally used to calculate the brightness temperature maps using the following 
equation  \citep{madau1997,Furlanetto2006}.
\begin{eqnarray}
\TB (\mathbf{x}) \!\!\!\! &\equiv& \!\!\!\! 27 ~ x_{\rm HI} (\mathbf{x}, z) [1+\delta_{\rm B}(\mathbf{x}, z)] \left(\frac{\OmegaB h^2}{0.023}\right) \nonumber\\
&\times& \!\!\!\!\left(\frac{0.15}{\Omegam h^2}\frac{1+z}{10}\right)^{1/2}\left[1-\frac{\TCMB(z)}{T_{\rm S}(\mathbf{x}, z)}\right]\,\rm{mK},
\nonumber \\
\label{brightnessT}
\end{eqnarray}
where  $\mathbf{x} = r_z \mathbf{\hat{n}}$ with $\hat{n}$ be the line of sight direction. The quantity $r_z$ is the comoving radial distance to redshift $z$. Here, the quantities $\delta_{\rm B}(z,\mathbf{x})$ and $x_{\rm HI}(z,\mathbf{x})$ represent baryonic density contrast and the neutral hydrogen fraction respectively. $\TCMB(z)=2.73\times (1+z)$ K denotes the CMBR temperature at redshift $z$. Note that we have not included the line of 
sight effects such as, effect of peculiar velocity \citep{mao12, ghara15a, 2016JApA...37...32M} and light-cone effect \citep{2012MNRAS.424.1877D, ghara15b, 2018IAUS..333...12D} in the brightness temperature maps.

\end{itemize}  

Before considering realistic models of reionization as we describe in the following sections, let us first see the impacts of these two different X-ray SEDs around isolated sources, one with a power-law spectrum and another that is derived from the composite spectrum presented here. The $\TB$ profiles around the sources crucially depend on the properties of the sources \citep{2016MNRAS.460..827G}.  The right panel of Figure \ref{image_p5spectrum} shows the brightness temperature profiles of the power-law and 
composite spectrum along with the stellar SED. We have assumed that the stellar mass of the galaxy is $10^8 ~\MSUN$ while the age of the source is 20 Myr. We have fixed the parameter $f_X$ to 0.05 while we choose $\alpha =1.5$ as the spectral index of the power-law SED. Clearly,  $\TB$ profiles around the sources for different cases are distinct from each other.  Different amount of soft X-rays in the SEDs of these sources result in different amount of heating in the IGM gas. The size of the emission region for the composite spectrum  is smaller compared to the power-law spectrum. This is due to presence of fewer number of soft X-ray photons in the composite spectrum. On the other hand, almost no emission region is present beyond the \HII ~region for the stellar spectrum. Since the Stellar SED does not contain any soft X-rays, we do not find any prominent emission region in the corresponding $\TB$ profile. As the absorbed HMXB spectrum from \citet{fragos2013} contains a negligible amount of soft X-ray photons, the heating is mostly dominated by hard X-rays. Due to the large mean free path, these hard X-rays travel long distances before getting absorbed after sufficient redshifting. Thus, the heating in the surrounding medium of the source is less in case of absorbed HMXB spectrum from \citet{fragos2013} compared to the other X-ray spectrum. On the other hand, the intrinsic HMXB spectrum of \citet{fragos2013} contains slightly more soft X-ray photons than the composite spectrum, but less compared to the power-law spectrum for $f_X=0.05$. Thus, the heating due to the intrinsic HMXB spectrum of \citet{fragos2013} is higher than that from the composite spectrum and lower compared to the power-law spectrum. Thus, the brightness temperature profile around the isolated source with unabsorbed spectrum remains in-between the 1D $\TB$ profiles of the composite and power-law spectrum.   The number of $\lya$ photons emitted by all the three sources are nearly the same, hence the $\TB$ profiles at far away distances (where the impact of X-rays is negligible) are identical in these cases. Although these profiles are different from each other, distinguishing these profiles in real observations may not be an easy task. In principle, these profiles can be distinctly determined in high-resolution images using long observation time  \citep{2017MNRAS.464.2234G}.

%%%%%%%%%%%%%%%%%%%%
\subsection{Impacts on reionization}
\label{sec:res}
\begin{figure}
\begin{center}
\includegraphics[scale=0.4]{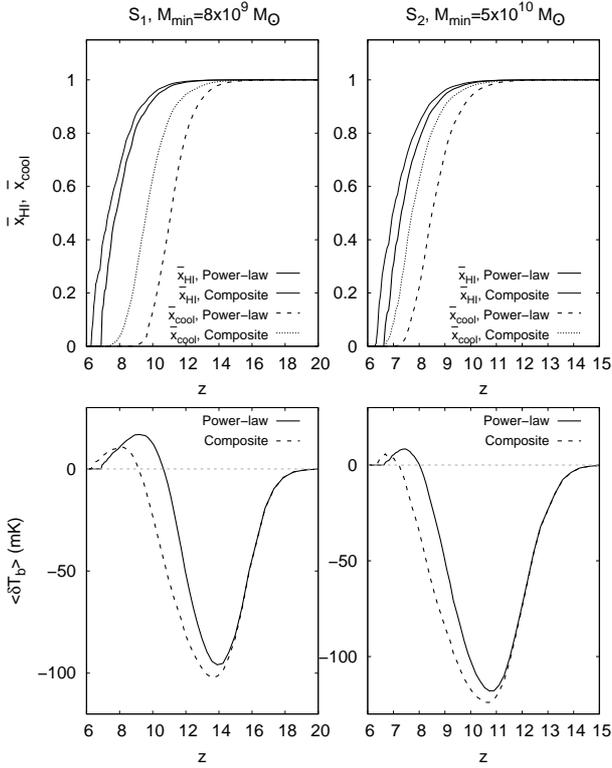}
    \caption{{\bf Upper panels:} The solid lines in the upper panels denote the evolution of the volume averaged ionization fraction of hydrogen as a function of     redshift. The dashed lines represent the redshift evolution of the ``cooled fraction'' of the IGM. Here, we define a region as ``cooled''     region if the kinetic temperature of the region is smaller than the CMBR temperature. The thick and thin dashed curves represent the power-law and 
    composite spectrum respectively.  {\bf Lower panels:} Redshift evolution of the volume averaged brightness temperature for different source models. The solid curve corresponds to the power-law model, while the dashed curve denotes the composite spectrum. In both source models, we fix the parameter $f_X$ as 0.05. The left and right panels represent scenarios $S_1$ and $S_2$ respectively which corresponds to  the minimum halo mass  $M_{\rm min}=8\times10^9 ~\MSUN$ and $5\times10^{10} ~\MSUN$ respectively.} 
   \label{image_p5ionfrac}
\end{center}
\end{figure}

Here, we have considered two different types of X-ray spectrum, in particular a power-law spectrum (corresponds to sources such as mini-QSO) and a composite spectrum of HMXBs and studied their impacts on the 21-cm signal from the cosmic dawn and EoR. In addition, we will consider two different scenarios with two different minimum halo mass ($M_{\rm min}$) that host sources of radiation. In particular, we choose $M_{\rm min}=8\times10^9 ~\MSUN$ and $5\times10^{10} ~\MSUN$ and denote these two scenarios by $S_1$ and $S_2$ respectively.  We choose $f_\star = 0.02$ (0.02) and $f_{\rm esc} = 0.15$ (0.6) for $S_1$ ($S_2$). This produces an ionization history which starts around redshift 20 (15) and ends around redshift 6.8 (6.6) and 6.2 (6.3) for the power-law and composite spectrum respectively in case of $S_1$ ($S_2$). (as shown by the solid curves in the upper panels of figure \ref{image_p5ionfrac}). The Thompson optical depths are $\tau=0.059$ (0.052) and 0.052 (0.05) for the power-law and composite spectrum respectively in scenario $S_1$ ($S_2$), which is consistent with the current observational constraint from $Planck$ mission \citep{2016A&A...596A.108P}. The dashed curves in the upper panels of  Figure \ref{image_p5ionfrac} show the evolution of the volume averaged ``cooled fraction ($\bar{x}_{\rm cool}$)'' for different source models. We define a region as ``cooled'' if the kinetic temperature of that region is smaller than the CMBR temperature. One can see that the IGM becomes heated quicker in the case of the power-law spectrum of the X-ray sources. This is due to the presence of more soft X-ray photons in the spectrum of power-law compared to the HMXBs composite spectrum.

\subsubsection{Global 21-cm signal}
\label{res_global}

The lower panels of Figure \ref{image_p5ionfrac} shows the evolution of the volume averaged $\TB$ as a function of the redshift for the different source models considered in this paper. The average $\TB$ starts to deviate from zero as soon as the first sources begin to form during the cosmic dawn. In subsequent time, the kinetic temperature of the IGM gets coupled to the spin temperature of \HI ~due to Wouthuysen-Field coupling \citep{wouth52, field58, hirata2006lya} by the $\lya$ photons from the first sources. This produces a strong absorption signal from \HI , as the IGM is not heated above CMBR temperature at that time. Eventually, the IGM becomes heated above $\TCMB$ in the presence of the X-ray sources, and the signal becomes emission signal. This produces a trough-like feature in the evolution curve of $<\TB>$ around redshift 14.5 for the power-law model and 14 for the composite spectrum model. Eventually, the signal 
vanishes as the IGM becomes highly ionized.  The evolution curve of  $<\TB>$ is, therefore, highly dependent on the nature of X-ray sources. These results are consistent with previous works such as \citet{ 2017MNRAS.472.1915C, 2018MNRAS.477.3217G, 2018MNRAS.478.5591M, 2019arXiv190202438F}. 

In case of $S_2$, the $\lya$ coupling and heating is delayed as only the massive halos with masses larger than $5\times10^{10}$ contribute to these processes. However, the choice of a larger escape fraction of the UV photons in $S_1$ scenario completes the reionization by redshift 6. As the $\lya$ coupling and X-ray heating is delayed, the brightness temperature vs redshift profile is also delayed as shown in the right bottom panel of Figure \ref{image_p5ionfrac}. In this case, the trough appears at redshift $\sim 11$. The minimum value of $<\TB>$ is slightly smaller in case of $S_2$ compared to $S_1$ due to the delayed heating in $S_2$.  

One can see that the signal amplitude is maximum during the first phase of reionization. Thus, many of the EoR experiments such as EDGES,  SARAS,  LEDA etc  which intent to detect the global signal, probe this redshift range. However, these low-frequency experiments are 
limited by the foregrounds, calibration, systematics of the instruments etc. The global absorption signal transforms into an emission signal as the IGM gets heated due to the presence of X-rays in this study. As the amount of heating is different in these model sources, the transition occurs at different redshifts. The transition redshifts are 11 and 10 for the power-law and composite spectrum driven reionization respectively. Also, the absorption signal is relatively stronger for the composite source model due to less heating compared to the power-law SED. Thus, it is expected that the signal is easier to detect for the composite spectrum in the global EoR experiments. As we will see later that the amplitude of the power spectrum for the reionization driven by the composite spectrum is smaller compared to the power-law spectrum, it will be hard to detect the composite spectrum driven reionization at the first phase of reionization by the statistical analysis, while the global experiments may be very useful.

The strongest absorption signal is $\sim -100$ mK in these models. Recently, \citet{EDGES2018} have reported a measurement of such redshift-brightness temperature profile of the average 21-cm signal around redshift $z\sim 17$ using the EDGES low-band observations. However, the measured amplitude of the signal is stronger by several factors than what our models predict. To explain such a strong signal as reported by EDGES, one has to consider either an excess cooling due to an unknown physical process such as baryon-dark matter interaction  \citep{2018Natur.555...71B, 2018PhRvL.121a1101F, 2018arXiv180210094M, PhysRevLett.121.011102} or an excess radio background \citep{2018ApJ...858L..17F, 2018arXiv180301815E, 2018PhLB..785..159F, 2019arXiv190202438F, 2019arXiv190208282B}. However, we have not considered any excess cooling beside the cooling due to the expansion of the Universe or any excess radio background while estimating the 21-cm signal in our study. The strongest absorption signal in Figure \ref{image_p5ionfrac} appears at redshift $\sim 14$. These absorption profiles are delayed as we have not considered contributions from halos with mass $\lesssim 8\times10^9 \MSUN$.

%%%%%%%%%%%%%%%

\subsubsection{Fluctuations in 21-cm signal}
\label{RES_fluc}

Next, we will study the implications of the composite spectrum on the fluctuation of the signal. Primarily, the first generation radio telescopes aim to measure the fluctuations in terms of the spherically averaged power spectrum $P(k)$ which can be written as
\begin{equation}
\langle \hat{\TB}(\mathrm{\bf k}) \hat{\TB}^{\star}(\mathbf{k'})\rangle = (2 \pi)^3 \delta_{\rm D}(\mathbf{k - k'}) P(k),
\label{ps}
\end{equation}
where $\hat{\TB}(\mathrm{\bf k})$ is the Fourier component of $\TB(\mathbf{x})$ at scale $k$. However, we will present our results in terms of the dimensionless power spectrum $\Delta^2(k)=k^3P(k)/2\pi^2$.

\begin{figure}
\begin{center}
\includegraphics[scale=0.41]{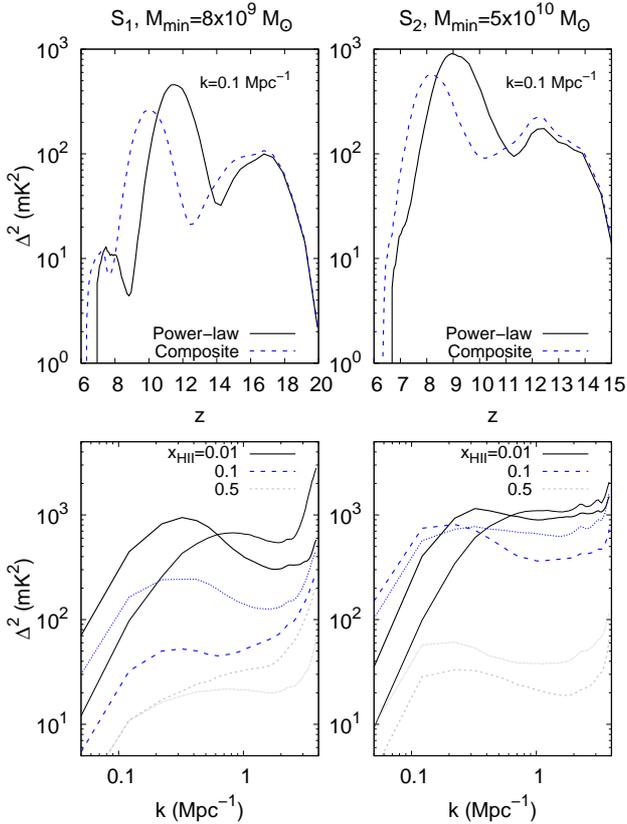}
    \caption{{ \bf Upper panels:} Redshift evolution of the large-scale power spectrum of $\TB$ for two different source model, dashed curve present the     composite spectrum while the  solid curve corresponds to power-law spectrum. The curves correspond to scale $0.1$ Mpc$^{-1}$. {\bf Lower panels:} The large-scale power spectrum of $\TB$ as a function of scale for the two different source models considered in this work, thin curves present 
    the composite spectrum while the  thick curves correspond to power-law spectrum. The left and right panels corresponds to scenarios $S_1$ and $S_2$ respectively.}
   \label{image_p5pszk}
\end{center}
\end{figure}

The upper panels of Figure \ref{image_p5pszk} shows the redshift evolution of the large-scale ($k = 0.1 ~\rm Mpc^{-1}$) power spectrum for the two different source models considered in the paper. The large-scale power spectrum shows three distinct peaks when plotted against redshift.  This is consistent with studies by \citet{ baek09, Mesinger2013, ghara15a, 2017MNRAS.464.3498F, 2018MNRAS.478.2193C}.  From high to low redshift, these peaks arise due to the fluctuation in $\lya$ 
flux at the cosmic dawn, heating inhomogeneity during the first phase of reionization and the ionization inhomogeneities in the later stages of reionization. These three characteristic peaks are prominent for the power-law spectrum and the composite spectrum. However, studies such as \citet{pacucci2014,fialkov2014} shows very distinct feature in the evolution of the large-scale power spectrum for the HMXBs spectrum. The HMXBs spectrum 
considered in these studies have very small number of soft X-ray photons compared to the hard X-ray photons and thus, the heating is dominated by the hard X-rays. As a result, the heating is homogeneous in the neutral IGM in these model, which lead to the disappearance of the heating peak around redshift $\sim 10$. However, our composite spectrum is basically a power-law but not very steep. Also, the effective number of soft X-rays are less  in the composite spectrum compared to the power-law scenario, as we have fixed the parameter $f_X=0.05$. Thus, the heating is inhomogeneous but late compared to the power-law spectrum. The global ionization is also delayed for the composite spectrum scenario due to the lack of soft ionizing photons in the SED. This late ionization for the composite spectrum scenario shifts the heating peak towards the lower redshift. On the other hand, less inhomogeneous heating for the composite SED results in decreasing the amplitude of the heating peak by factor 2 compared to the power-law SED in case of $S_1$.   The $\lya$ fluctuation peak and heating peak appear at lower redshifts in case of $S_2$ compared to $S_1$ as $\lya$ coupling and X-ray heating is delayed due to the absence of halos with mass less than $5\times 10^{10} \MSUN$ in $S_2$. One can also easily notice that the amplitudes of the $\lya$ fluctuation and heating peaks are larger in case of $S_2$ than those of $S_1$. These results are consistent with studies such as \citet{das2017}.

It is not quite clear whether one should compare the scale dependence of the power spectrum at certain redshifts or for certain ionization fraction. Here, we compare the power spectrum from these two source models at fixed ionization fractions. The scale dependent of the power spectrum is shown in the lower panels of the Figure \ref{image_p5pszk} for different ionization fractions 0.01, 0.1 and 0.5 from the two source models. The redshifts are 
(11.73, 9.6, 7.8) and (11.3, 9.2, 7.3) for the power-law and composite spectrum respectively. We see a bump around $k \sim 0.3 ~\rm Mpc^{-1}$ at redshift 11.73 curve for the power-law scenario. This corresponds to the characteristic size  of the heated bubble at that redshift, as the $\lya$ coupling is saturated by then. In this case, $\rm R_{heat} = 2.46/k_{peak} \sim 8 ~cMpc$ \citep{friedrich11}. However, this peak probably arises due 
to the fact that there are no small mass halos in our simulation and thus the heated bubbles are relatively rare and with less overlap. With small mass haloes included the peak should move towards small scales \citep{ghara15a}.  Another fact to notice is that the peak moves towards small scales $(\rm k_{peak} \sim 0.6)$, which correspond to a smaller characteristic size of the  heated bubble for the composite spectrum scenario. This is expected as we have seen the size of the emission regions is less for the composite spectrum case (see Figure \ref{image_p5spectrum}). The peaky feature of the power spectrum as a function of scales decreases as reionization progresses and disappears as soon as the IGM gets sufficiently 
heated. 

\section{Summary and discussion}
\label{conc}
In this paper, the average spectra of galactic X-ray binaries are investigated using the long term monitoring data of {\it MAXI}--GSC. This sample of 17 HMXBs and 40 LMXBs, with sufficient statistics to extract spectra with $\sim$ 5.2 years of {\it MAXI}--GSC data, consists of various sub-classes of X-ray binaries. These average spectra are utilised to construct the composite spectrum of X-ray binaries. We use the composite spectra of galactic HMXBs to investigate the contribution of X-ray binaries heating in the redshifted 21-cm signal. Similar studies such as \citet{pacucci2014, fialkov2014} using semi-numerical methods have been done before. Here, we have used the observationally motivated spectrum for the HMXBs and used 1D radiative transfer method to generate the signal. The main outcomes can be summarized as follows.

\begin{itemize}
\item The composite spectrum of the HMXBs can be represented as a piece-wise power-law with three components in the energy range $3\hbox{--}20$ keV. Having extrapolated the power-law dependence towards soft X-ray part (below 3 keV), we have studied the different signatures of this composite spectrum and simple power-law spectrum keeping the fraction of the X-ray and UV luminosity fixed. 

\item $\TB$ profiles around isolated source shows different feature for the composite spectrum compared to the power-law sources. Since the number of the soft X-ray photons is less in the composite spectrum, the emission region around the central \HII ~region is smaller. This makes the profile smoother than the profile around a stellar type source and sharper than the profile around the power-law spectrum. Distinguishing these profiles need 
high-resolution imaging and a long observation time, and thus may not be an easy task. The ionization front around the isolated sources is smaller for the composite spectrum, which results in a delayed reionization. 

\item In the presence of X-rays, the absorption signal transforms into emission after a certain stage of reionization. For the composite spectrum, the transition occurs much later than the power-law spectrum for the same $f_X$. The amplitude of the dip in $<\TB>$ is larger for the composite spectrum than the power-law driven scenario, which implies that it will be easier to detect the signal in the global 21-cm experiments.    

\item In the presence of soft X-rays in the power-law and composite spectrum, the heating is patchy. This gives rise to a distinct peak in the evolution curve of the large-scale power spectrum of the signal. Since the composite spectrum  of the HMXBs is less steep compared to the power-law spectrum, the heating of the IGM is less patchy. Also, fewer soft X-ray photons in the composite spectrum results in late heating which causes a shift of the 
heating peak of the large-scale power spectrum when plotted against redshift towards lower redshift side. In addition, the amplitude of the heating peak decreases for the composite spectrum.  These results are consistent with those of earlier works by \citet{Mesinger2013, pacucci2014}.

\item  For a larger value of the minimum mass of halos ($M_{\rm min}$) that contains sources, the 21-cm signal is delayed due to late $\lya$ coupling and X-ray heating. The minimum value of the average brightness temperature appears late and smaller for a larger value of $M_{\rm min}$. On the other hand, the maximum value of the large-scale power spectrum enhances for a larger value of $M_{\rm min}$.  
\end{itemize}

While we have studied the difference between the implications of the composite spectrum of the HMXBs and simple power-law spectrum, one should also check the detectability of these source models with present and future radio observations in the presence of system noise, foregrounds etc.

\section*{Acknowledgement}
We thank the anonymous referee for insightful comments and suggestions.
This research has made use of MAXI data provided by RIKEN, JAXA and the MAXI team. RG would like to thank Samir Choudhuri for his assistance.

\bibliography{bibtex}

\begin{thebibliography}{}
\makeatletter
\relax
\def\mn@urlcharsother{\let\do\@makeother \do\$\do\&\do\#\do\^\do\_\do\%\do\~}
\def\mn@doi{\begingroup\mn@urlcharsother \@ifnextchar [ {\mn@doi@}
  {\mn@doi@[]}}
\def\mn@doi@[#1]#2{\def\@tempa{#1}\ifx\@tempa\@empty \href
  {http://dx.doi.org/#2} {doi:#2}\else \href {http://dx.doi.org/#2} {#1}\fi
  \endgroup}
\def\mn@eprint#1#2{\mn@eprint@#1:#2::\@nil}
\def\mn@eprint@arXiv#1{\href {http://arxiv.org/abs/#1} {{\tt arXiv:#1}}}
\def\mn@eprint@dblp#1{\href {http://dblp.uni-trier.de/rec/bibtex/#1.xml}
  {dblp:#1}}
\def\mn@eprint@#1:#2:#3:#4\@nil{\def\@tempa {#1}\def\@tempb {#2}\def\@tempc
  {#3}\ifx \@tempc \@empty \let \@tempc \@tempb \let \@tempb \@tempa \fi \ifx
  \@tempb \@empty \def\@tempb {arXiv}\fi \@ifundefined
  {mn@eprint@\@tempb}{\@tempb:\@tempc}{\expandafter \expandafter \csname
  mn@eprint@\@tempb\endcsname \expandafter{\@tempc}}}

\bibitem[\protect\citeauthoryear{{Arnaud}}{{Arnaud}}{1996}]{arnaud1996}
{Arnaud} K.~A.,  1996, in {Jacoby} G.~H.,  {Barnes} J.,  eds,  Astronomical
  Society of the Pacific Conference Series Vol. 101, Astronomical Data Analysis
  Software and Systems V. p.~17

\bibitem[\protect\citeauthoryear{{Baek}, {Di Matteo}, {Semelin}, {Combes}  \&
  {Revaz}}{{Baek} et~al.}{2009}]{baek09}
{Baek} S.,  {Di Matteo} P.,  {Semelin} B.,  {Combes} F.,   {Revaz} Y.,  2009,
  \mn@doi [\aap] {10.1051/0004-6361:200810757}, \href
  {http://adsabs.harvard.edu/abs/2009A%26A...495..389B} {495, 389}

\bibitem[\protect\citeauthoryear{{Barkana}}{{Barkana}}{2018}]{2018Natur.555...71B}
{Barkana} R.,  2018, \mn@doi [\nat] {10.1038/nature25791}, \href
  {http://adsabs.harvard.edu/abs/2018Natur.555...71B} {555, 71}

\bibitem[\protect\citeauthoryear{Berlin, Hooper, Krnjaic  \& McDermott}{Berlin
  et~al.}{2018}]{PhysRevLett.121.011102}
Berlin A.,  Hooper D.,  Krnjaic G.,   McDermott S.~D.,  2018, \mn@doi [Phys.
  Rev. Lett.] {10.1103/PhysRevLett.121.011102}, 121, 011102

\bibitem[\protect\citeauthoryear{{Bowman} \& {Rogers}}{{Bowman} \&
  {Rogers}}{2010}]{2010Natur.468..796B}
{Bowman} J.~D.,  {Rogers} A.~E.~E.,  2010, \mn@doi [\nat]
  {10.1038/nature09601}, \href
  {http://adsabs.harvard.edu/abs/2010Natur.468..796B} {468, 796}

\bibitem[\protect\citeauthoryear{{Bowman} et~al.,}{{Bowman}
  et~al.}{2013}]{bowman13}
{Bowman} J.~D.,  et~al., 2013, \mn@doi [\pasa] {10.1017/pas.2013.009}, \href
  {http://adsabs.harvard.edu/abs/2013PASA...30...31B} {30, e031}

\bibitem[\protect\citeauthoryear{{Bowman}, {Rogers}, {Monsalve}, {Mozdzen}  \&
  {Mahesh}}{{Bowman} et~al.}{2018}]{EDGES2018}
{Bowman} J.~D.,  {Rogers} A.~E.~E.,  {Monsalve} R.~A.,  {Mozdzen} T.~J.,
  {Mahesh} N.,  2018, \mn@doi [\nat] {10.1038/nature25792}, \href
  {http://adsabs.harvard.edu/abs/2018Natur.555...67B} {555, 67}

\bibitem[\protect\citeauthoryear{{Brandenberger}, {Cyr}  \&
  {Shi}}{{Brandenberger} et~al.}{2019}]{2019arXiv190208282B}
{Brandenberger} R.,  {Cyr} B.,   {Shi} R.,  2019, arXiv e-prints, \href
  {https://ui.adsabs.harvard.edu/\#abs/2019arXiv190208282B} {p.
  arXiv:1902.08282}

\bibitem[\protect\citeauthoryear{{Cohen}, {Fialkov}, {Barkana}  \&
  {Lotem}}{{Cohen} et~al.}{2017}]{2017MNRAS.472.1915C}
{Cohen} A.,  {Fialkov} A.,  {Barkana} R.,   {Lotem} M.,  2017, \mn@doi [\mnras]
  {10.1093/mnras/stx2065}, \href
  {https://ui.adsabs.harvard.edu/\#abs/2017MNRAS.472.1915C} {472, 1915}

\bibitem[\protect\citeauthoryear{{Cohen}, {Fialkov}  \& {Barkana}}{{Cohen}
  et~al.}{2018}]{2018MNRAS.478.2193C}
{Cohen} A.,  {Fialkov} A.,   {Barkana} R.,  2018, \mn@doi [\mnras]
  {10.1093/mnras/sty1094}, \href
  {https://ui.adsabs.harvard.edu/\#abs/2018MNRAS.478.2193C} {478, 2193}

\bibitem[\protect\citeauthoryear{{Corbet} \& {Krimm}}{{Corbet} \&
  {Krimm}}{2013}]{corbet2013}
{Corbet} R.~H.~D.,  {Krimm} H.~A.,  2013, \mn@doi [\apj]
  {10.1088/0004-637X/778/1/45}, \href
  {http://adsabs.harvard.edu/abs/2013ApJ...778...45C} {778, 45}

\bibitem[\protect\citeauthoryear{{Corbet} et~al.,}{{Corbet}
  et~al.}{2007}]{corbet2007}
{Corbet} R.,  et~al., 2007, \mn@doi [Progress of Theoretical Physics
  Supplement] {10.1143/PTPS.169.200}, \href
  {http://adsabs.harvard.edu/abs/2007PThPS.169..200C} {169, 200}

\bibitem[\protect\citeauthoryear{{Das}, {Mesinger}, {Pallottini}, {Ferrara}  \&
  {Wise}}{{Das} et~al.}{2017}]{das2017}
{Das} A.,  {Mesinger} A.,  {Pallottini} A.,  {Ferrara} A.,   {Wise} J.~H.,
  2017, \mn@doi [\mnras] {10.1093/mnras/stx943}, \href
  {https://ui.adsabs.harvard.edu/#abs/2017MNRAS.469.1166D} {469, 1166}

\bibitem[\protect\citeauthoryear{{Datta}, {Mellema}, {Mao}, {Iliev}, {Shapiro}
  \& {Ahn}}{{Datta} et~al.}{2012}]{2012MNRAS.424.1877D}
{Datta} K.~K.,  {Mellema} G.,  {Mao} Y.,  {Iliev} I.~T.,  {Shapiro} P.~R.,
  {Ahn} K.,  2012, \mn@doi [\mnras] {10.1111/j.1365-2966.2012.21293.x}, \href
  {http://adsabs.harvard.edu/abs/2012MNRAS.424.1877D} {424, 1877}

\bibitem[\protect\citeauthoryear{{Datta}, {Mondal}, {Ghara}, {Bharadwaj}  \&
  {Choudhury}}{{Datta} et~al.}{2018}]{2018IAUS..333...12D}
{Datta} K.~K.,  {Mondal} R.,  {Ghara} R.,  {Bharadwaj} S.,   {Choudhury} T.~R.,
   2018, in {Jeli{\'c}} V.,  {van der Hulst} T.,  eds,  IAU Symposium Vol. 333,
  Peering towards Cosmic Dawn. pp 12--17, \mn@doi{10.1017/S1743921318000637}

\bibitem[\protect\citeauthoryear{{DeBoer} et~al.,}{{DeBoer}
  et~al.}{2017}]{2017PASP..129d5001D}
{DeBoer} D.~R.,  et~al., 2017, \mn@doi [Publications of the Astronomical
  Society of the Pacific] {10.1088/1538-3873/129/974/045001}, \href
  {https://ui.adsabs.harvard.edu/\#abs/2017PASP..129d5001D} {129, 045001}

\bibitem[\protect\citeauthoryear{{Doroshenko}, {Santangelo}, {Nakahira},
  {Mihara}, {Sugizaki}, {Matsuoka}, {Nakajima}  \& {Makishima}}{{Doroshenko}
  et~al.}{2013}]{doroshenkov2013}
{Doroshenko} V.,  {Santangelo} A.,  {Nakahira} S.,  {Mihara} T.,  {Sugizaki}
  M.,  {Matsuoka} M.,  {Nakajima} M.,   {Makishima} K.,  2013, \mn@doi [\aap]
  {10.1051/0004-6361/201321305}, \href
  {http://adsabs.harvard.edu/abs/2013A%26A...554A..37D} {554, A37}

\bibitem[\protect\citeauthoryear{{Ewall-Wice}, {Chang}, {Lazio}, {Dor{\'e}},
  {Seiffert}  \& {Monsalve}}{{Ewall-Wice} et~al.}{2018}]{2018arXiv180301815E}
{Ewall-Wice} A.,  {Chang} T.-C.,  {Lazio} J.,  {Dor{\'e}} O.,  {Seiffert} M.,
  {Monsalve} R.~A.,  2018, preprint, \href
  {http://adsabs.harvard.edu/abs/2018arXiv180301815E} {} (\mn@eprint {arXiv}
  {1803.01815})

\bibitem[\protect\citeauthoryear{{Fabbiano}}{{Fabbiano}}{2006}]{fabbiano2006}
{Fabbiano} G.,  2006, \mn@doi [Annual Review of Astronomy and Astrophysics]
  {10.1146/annurev.astro.44.051905.092519}, \href
  {https://ui.adsabs.harvard.edu/#abs/2006ARA&A..44..323F} {44, 323}

\bibitem[\protect\citeauthoryear{{Feng} \& {Holder}}{{Feng} \&
  {Holder}}{2018}]{2018ApJ...858L..17F}
{Feng} C.,  {Holder} G.,  2018, \mn@doi [\apjl] {10.3847/2041-8213/aac0fe},
  \href {http://adsabs.harvard.edu/abs/2018ApJ...858L..17F} {858, L17}

\bibitem[\protect\citeauthoryear{{Fialkov} \& {Barkana}}{{Fialkov} \&
  {Barkana}}{2019}]{2019arXiv190202438F}
{Fialkov} A.,  {Barkana} R.,  2019, arXiv e-prints, \href
  {https://ui.adsabs.harvard.edu/\#abs/2019arXiv190202438F} {p.
  arXiv:1902.02438}

\bibitem[\protect\citeauthoryear{{Fialkov}, {Barkana}  \& {Visbal}}{{Fialkov}
  et~al.}{2014}]{fialkov2014}
{Fialkov} A.,  {Barkana} R.,   {Visbal} E.,  2014, \mn@doi [\nat]
  {10.1038/nature12999}, \href
  {http://adsabs.harvard.edu/abs/2014Natur.506..197F} {506, 197}

\bibitem[\protect\citeauthoryear{{Fialkov}, {Cohen}, {Barkana}  \&
  {Silk}}{{Fialkov} et~al.}{2017}]{2017MNRAS.464.3498F}
{Fialkov} A.,  {Cohen} A.,  {Barkana} R.,   {Silk} J.,  2017, \mn@doi [\mnras]
  {10.1093/mnras/stw2540}, \href
  {https://ui.adsabs.harvard.edu/\#abs/2017MNRAS.464.3498F} {464, 3498}

\bibitem[\protect\citeauthoryear{{Fialkov}, {Barkana}  \& {Cohen}}{{Fialkov}
  et~al.}{2018}]{2018PhRvL.121a1101F}
{Fialkov} A.,  {Barkana} R.,   {Cohen} A.,  2018, \mn@doi [Physical Review
  Letters] {10.1103/PhysRevLett.121.011101}, \href
  {http://adsabs.harvard.edu/abs/2018PhRvL.121a1101F} {121, 011101}

\bibitem[\protect\citeauthoryear{{Field}}{{Field}}{1958}]{field58}
{Field} G.~B.,  1958, \mn@doi [Proceedings of the IRE]
  {10.1109/JRPROC.1958.286741}, \href
  {http://adsabs.harvard.edu/abs/1958PIRE...46..240F} {46, 240}

\bibitem[\protect\citeauthoryear{{Fioc} \& {Rocca-Volmerange}}{{Fioc} \&
  {Rocca-Volmerange}}{1997}]{Fioc97}
{Fioc} M.,  {Rocca-Volmerange} B.,  1997, \aap, \href
  {http://adsabs.harvard.edu/abs/1997A%26A...326..950F} {326, 950}

\bibitem[\protect\citeauthoryear{{Fragos}, {Lehmer}, {Naoz}, {Zezas}  \&
  {Basu-Zych}}{{Fragos} et~al.}{2013}]{fragos2013}
{Fragos} T.,  {Lehmer} B.~D.,  {Naoz} S.,  {Zezas} A.,   {Basu-Zych} A.,  2013,
  \mn@doi [\apjl] {10.1088/2041-8205/776/2/L31}, \href
  {http://adsabs.harvard.edu/abs/2013ApJ...776L..31F} {776, L31}

\bibitem[\protect\citeauthoryear{{Fraser} et~al.,}{{Fraser}
  et~al.}{2018}]{2018PhLB..785..159F}
{Fraser} S.,  et~al., 2018, \mn@doi [Physics Letters B]
  {10.1016/j.physletb.2018.08.035}, \href
  {http://adsabs.harvard.edu/abs/2018PhLB..785..159F} {785, 159}

\bibitem[\protect\citeauthoryear{{Fridriksson} et~al.,}{{Fridriksson}
  et~al.}{2011}]{fridriksson2011}
{Fridriksson} J.~K.,  et~al., 2011, \mn@doi [\apj]
  {10.1088/0004-637X/736/2/162}, \href
  {https://ui.adsabs.harvard.edu/\#abs/2011ApJ...736..162F} {736, 162}

\bibitem[\protect\citeauthoryear{{Friedrich}, {Mellema}, {Alvarez}, {Shapiro}
  \& {Iliev}}{{Friedrich} et~al.}{2011}]{friedrich11}
{Friedrich} M.~M.,  {Mellema} G.,  {Alvarez} M.~A.,  {Shapiro} P.~R.,   {Iliev}
  I.~T.,  2011, \mn@doi [\mnras] {10.1111/j.1365-2966.2011.18219.x}, \href
  {http://adsabs.harvard.edu/abs/2011MNRAS.413.1353F} {413, 1353}

\bibitem[\protect\citeauthoryear{{Furlanetto}, {Oh}  \& {Briggs}}{{Furlanetto}
  et~al.}{2006}]{Furlanetto2006}
{Furlanetto} S.~R.,  {Oh} S.~P.,   {Briggs} F.~H.,  2006, \mn@doi [\physrep]
  {10.1016/j.physrep.2006.08.002}, \href
  {http://adsabs.harvard.edu/abs/2006PhR...433..181F} {433, 181}

\bibitem[\protect\citeauthoryear{{Ghara}, {Choudhury}  \& {Datta}}{{Ghara}
  et~al.}{2015a}]{ghara15a}
{Ghara} R.,  {Choudhury} T.~R.,   {Datta} K.~K.,  2015a, \mn@doi [\mnras]
  {10.1093/mnras/stu2512}, \href
  {http://adsabs.harvard.edu/abs/2015MNRAS.447.1806G} {447, 1806}

\bibitem[\protect\citeauthoryear{{Ghara}, {Datta}  \& {Choudhury}}{{Ghara}
  et~al.}{2015b}]{ghara15b}
{Ghara} R.,  {Datta} K.~K.,   {Choudhury} T.~R.,  2015b, \mn@doi [\mnras]
  {10.1093/mnras/stv1855}, \href
  {http://adsabs.harvard.edu/abs/2015MNRAS.453.3143G} {453, 3143}

\bibitem[\protect\citeauthoryear{{Ghara}, {Choudhury}  \& {Datta}}{{Ghara}
  et~al.}{2016}]{2016MNRAS.460..827G}
{Ghara} R.,  {Choudhury} T.~R.,   {Datta} K.~K.,  2016, \mn@doi [\mnras]
  {10.1093/mnras/stw953}, \href
  {http://adsabs.harvard.edu/abs/2016MNRAS.460..827G} {460, 827}

\bibitem[\protect\citeauthoryear{{Ghara}, {Choudhury}, {Datta}  \&
  {Choudhuri}}{{Ghara} et~al.}{2017}]{2017MNRAS.464.2234G}
{Ghara} R.,  {Choudhury} T.~R.,  {Datta} K.~K.,   {Choudhuri} S.,  2017,
  \mn@doi [\mnras] {10.1093/mnras/stw2494}, \href
  {http://adsabs.harvard.edu/abs/2017MNRAS.464.2234G} {464, 2234}

\bibitem[\protect\citeauthoryear{{Ghara}, {Mellema}, {Giri}, {Choudhury},
  {Datta}  \& {Majumdar}}{{Ghara} et~al.}{2018}]{2018MNRAS.476.1741G}
{Ghara} R.,  {Mellema} G.,  {Giri} S.~K.,  {Choudhury} T.~R.,  {Datta} K.~K.,
  {Majumdar} S.,  2018, \mn@doi [\mnras] {10.1093/mnras/sty314}, \href
  {http://adsabs.harvard.edu/abs/2018MNRAS.476.1741G} {476, 1741}

\bibitem[\protect\citeauthoryear{{Giacconi}, {Gursky}, {Paolini}  \&
  {Rossi}}{{Giacconi} et~al.}{1962}]{giacconi1962}
{Giacconi} R.,  {Gursky} H.,  {Paolini} F.~R.,   {Rossi} B.~B.,  1962, \mn@doi
  [Physical Review Letters] {10.1103/PhysRevLett.9.439}, \href
  {http://adsabs.harvard.edu/abs/1962PhRvL...9..439G} {9, 439}

\bibitem[\protect\citeauthoryear{{Greig} \& {Mesinger}}{{Greig} \&
  {Mesinger}}{2018}]{2018MNRAS.477.3217G}
{Greig} B.,  {Mesinger} A.,  2018, \mn@doi [\mnras] {10.1093/mnras/sty796},
  \href {https://ui.adsabs.harvard.edu/\#abs/2018MNRAS.477.3217G} {477, 3217}

\bibitem[\protect\citeauthoryear{{Haberl} \& {Sturm}}{{Haberl} \&
  {Sturm}}{2016}]{haberl2016}
{Haberl} F.,  {Sturm} R.,  2016, \mn@doi [\aap] {10.1051/0004-6361/201527326},
  \href {https://ui.adsabs.harvard.edu/abs/2016A&A...586A..81H} {586, A81}

\bibitem[\protect\citeauthoryear{{Harnois-D{\'e}raps}, {Pen}, {Iliev}, {Merz},
  {Emberson}  \& {Desjacques}}{{Harnois-D{\'e}raps} et~al.}{2013}]{Harnois12}
{Harnois-D{\'e}raps} J.,  {Pen} U.-L.,  {Iliev} I.~T.,  {Merz} H.,  {Emberson}
  J.~D.,   {Desjacques} V.,  2013, \mn@doi [\mnras] {10.1093/mnras/stt1591},
  \href {http://adsabs.harvard.edu/abs/2013MNRAS.436..540H} {436, 540}

\bibitem[\protect\citeauthoryear{{Hasinger} \& {van der Klis}}{{Hasinger} \&
  {van der Klis}}{1989}]{hasinger1989}
{Hasinger} G.,  {van der Klis} M.,  1989, \aap, \href
  {https://ui.adsabs.harvard.edu/\#abs/1989A&A...225...79H} {225, 79}

\bibitem[\protect\citeauthoryear{{Hirata}}{{Hirata}}{2006}]{hirata2006lya}
{Hirata} C.~M.,  2006, \mn@doi [\mnras] {10.1111/j.1365-2966.2005.09949.x},
  \href {http://adsabs.harvard.edu/abs/2006MNRAS.367..259H} {367, 259}

\bibitem[\protect\citeauthoryear{{Islam} \& {Paul}}{{Islam} \&
  {Paul}}{2014}]{islam2014}
{Islam} N.,  {Paul} B.,  2014, \mn@doi [\mnras] {10.1093/mnras/stu756}, \href
  {http://adsabs.harvard.edu/abs/2014MNRAS.441.2539I} {441, 2539}

\bibitem[\protect\citeauthoryear{{Islam} \& {Paul}}{{Islam} \&
  {Paul}}{2016}]{islam2016}
{Islam} N.,  {Paul} B.,  2016, \mn@doi [\na] {10.1016/j.newast.2016.02.008},
  \href {https://ui.adsabs.harvard.edu/#abs/2016NewA...47...81I} {47, 81}

\bibitem[\protect\citeauthoryear{{Islam}, {Maitra}, {Pradhan}  \&
  {Paul}}{{Islam} et~al.}{2015}]{islam2015}
{Islam} N.,  {Maitra} C.,  {Pradhan} P.,   {Paul} B.,  2015, \mn@doi [\mnras]
  {10.1093/mnras/stu2395}, \href
  {https://ui.adsabs.harvard.edu/#abs/2015MNRAS.446.4148I} {446, 4148}

\bibitem[\protect\citeauthoryear{{Krimm} et~al.,}{{Krimm}
  et~al.}{2013}]{krimm2013}
{Krimm} H.~A.,  et~al., 2013, \mn@doi [The Astrophysical Journal Supplement
  Series] {10.1088/0067-0049/209/1/14}, \href
  {https://ui.adsabs.harvard.edu/\#abs/2013ApJS..209...14K} {209, 14}

\bibitem[\protect\citeauthoryear{{Lehmer} et~al.,}{{Lehmer}
  et~al.}{2015}]{lehmer2015}
{Lehmer} B.~D.,  et~al., 2015, \mn@doi [\apj] {10.1088/0004-637X/806/1/126},
  \href {http://adsabs.harvard.edu/abs/2015ApJ...806..126L} {806, 126}

\bibitem[\protect\citeauthoryear{{Lin}, {Remillard}  \& {Homan}}{{Lin}
  et~al.}{2007}]{lin2007}
{Lin} D.,  {Remillard} R.~A.,   {Homan} J.,  2007, \mn@doi [\apj]
  {10.1086/521181}, \href
  {https://ui.adsabs.harvard.edu/\#abs/2007ApJ...667.1073L} {667, 1073}

\bibitem[\protect\citeauthoryear{{Madau}, {Meiksin}  \& {Rees}}{{Madau}
  et~al.}{1997}]{madau1997}
{Madau} P.,  {Meiksin} A.,   {Rees} M.~J.,  1997, \mn@doi [\apj]
  {10.1086/303549}, \href {http://adsabs.harvard.edu/abs/1997ApJ...475..429M}
  {475, 429}

\bibitem[\protect\citeauthoryear{{Maitra} \& {Paul}}{{Maitra} \&
  {Paul}}{2013}]{maitra2013}
{Maitra} C.,  {Paul} B.,  2013, \mn@doi [\apj] {10.1088/0004-637X/771/2/96},
  \href {https://ui.adsabs.harvard.edu/#abs/2013ApJ...771...96M} {771, 96}

\bibitem[\protect\citeauthoryear{{Majumdar}, {Datta}, {Ghara}, {Mondal},
  {Choudhury}, {Bharadwaj}, {Ali}  \& {Datta}}{{Majumdar}
  et~al.}{2016}]{2016JApA...37...32M}
{Majumdar} S.,  {Datta} K.~K.,  {Ghara} R.,  {Mondal} R.,  {Choudhury} T.~R.,
  {Bharadwaj} S.,  {Ali} S.~S.,   {Datta} A.,  2016, \mn@doi [Journal of
  Astrophysics and Astronomy] {10.1007/s12036-016-9402-0}, \href
  {http://adsabs.harvard.edu/abs/2016JApA...37...32M} {37, 32}

\bibitem[\protect\citeauthoryear{{Mao}, {Shapiro}, {Mellema}, {Iliev}, {Koda}
  \& {Ahn}}{{Mao} et~al.}{2012}]{mao12}
{Mao} Y.,  {Shapiro} P.~R.,  {Mellema} G.,  {Iliev} I.~T.,  {Koda} J.,   {Ahn}
  K.,  2012, \mn@doi [\mnras] {10.1111/j.1365-2966.2012.20471.x}, \href
  {http://adsabs.harvard.edu/abs/2012MNRAS.422..926M} {422, 926}

\bibitem[\protect\citeauthoryear{{Matsuoka} et~al.,}{{Matsuoka}
  et~al.}{2009}]{matsuoka2009}
{Matsuoka} M.,  et~al., 2009, \pasj, \href
  {http://adsabs.harvard.edu/abs/2009PASJ...61..999M} {61, 999}

\bibitem[\protect\citeauthoryear{{McQuinn}}{{McQuinn}}{2012}]{2012MNRAS.426.1349M}
{McQuinn} M.,  2012, \mn@doi [\mnras] {10.1111/j.1365-2966.2012.21792.x}, \href
  {http://adsabs.harvard.edu/abs/2012MNRAS.426.1349M} {426, 1349}

\bibitem[\protect\citeauthoryear{{Mesinger}, {Ferrara}  \&
  {Spiegel}}{{Mesinger} et~al.}{2013}]{Mesinger2013}
{Mesinger} A.,  {Ferrara} A.,   {Spiegel} D.~S.,  2013, \mn@doi [\mnras]
  {10.1093/mnras/stt198}, \href
  {http://adsabs.harvard.edu/abs/2013MNRAS.431..621M} {431, 621}

\bibitem[\protect\citeauthoryear{{Mihara} et~al.,}{{Mihara}
  et~al.}{2011}]{mihara2011}
{Mihara} T.,  et~al., 2011, \pasj, \href
  {http://adsabs.harvard.edu/abs/2011PASJ...63S.623M} {63, 623}

\bibitem[\protect\citeauthoryear{{Mineo}, {Gilfanov}  \& {Sunyaev}}{{Mineo}
  et~al.}{2012}]{mineo2012}
{Mineo} S.,  {Gilfanov} M.,   {Sunyaev} R.,  2012, \mn@doi [\mnras]
  {10.1111/j.1365-2966.2011.19862.x}, \href
  {https://ui.adsabs.harvard.edu/#abs/2012MNRAS.419.2095M} {419, 2095}

\bibitem[\protect\citeauthoryear{{Mirabel}, {Dijkstra}, {Laurent}, {Loeb}  \&
  {Pritchard}}{{Mirabel} et~al.}{2011}]{mirabel2011}
{Mirabel} I.~F.,  {Dijkstra} M.,  {Laurent} P.,  {Loeb} A.,   {Pritchard}
  J.~R.,  2011, \mn@doi [\aap] {10.1051/0004-6361/201016357}, \href
  {http://adsabs.harvard.edu/abs/2011A%26A...528A.149M} {528, A149}

\bibitem[\protect\citeauthoryear{{Mirocha}, {Mebane}, {Furlanetto}, {Singal}
  \& {Trinh}}{{Mirocha} et~al.}{2018}]{2018MNRAS.478.5591M}
{Mirocha} J.,  {Mebane} R.~H.,  {Furlanetto} S.~R.,  {Singal} K.,   {Trinh} D.,
   2018, \mn@doi [\mnras] {10.1093/mnras/sty1388}, \href
  {https://ui.adsabs.harvard.edu/\#abs/2018MNRAS.478.5591M} {478, 5591}

\bibitem[\protect\citeauthoryear{{Mitsuda} et~al.,}{{Mitsuda}
  et~al.}{1984}]{mitsuda1984}
{Mitsuda} K.,  et~al., 1984, \pasj, \href
  {http://adsabs.harvard.edu/abs/1984PASJ...36..741M} {36, 741}

\bibitem[\protect\citeauthoryear{{Monsalve}, {Rogers}, {Bowman}  \&
  {Mozdzen}}{{Monsalve} et~al.}{2017}]{monsalve2017}
{Monsalve} R.~A.,  {Rogers} A. E.~E.,  {Bowman} J.~D.,   {Mozdzen} T.~J.,
  2017, \mn@doi [\apj] {10.3847/1538-4357/835/1/49}, \href
  {https://ui.adsabs.harvard.edu/\#abs/2017ApJ...835...49M} {835, 49}

\bibitem[\protect\citeauthoryear{{Mu{\~n}oz} \& {Loeb}}{{Mu{\~n}oz} \&
  {Loeb}}{2018}]{2018arXiv180210094M}
{Mu{\~n}oz} J.~B.,  {Loeb} A.,  2018, arXiv e-prints, \href
  {https://ui.adsabs.harvard.edu/\#abs/2018arXiv180210094M} {p.
  arXiv:1802.10094}

\bibitem[\protect\citeauthoryear{{Nagase}}{{Nagase}}{1989}]{nagase1989}
{Nagase} F.,  1989, \pasj, \href
  {http://adsabs.harvard.edu/abs/1989PASJ...41....1N} {41, 1}

\bibitem[\protect\citeauthoryear{{Nakahira} et~al.,}{{Nakahira}
  et~al.}{2012}]{nakahira2012}
{Nakahira} S.,  et~al., 2012, \pasj, \href
  {http://adsabs.harvard.edu/abs/2012PASJ...64...13N} {64, 13}

\bibitem[\protect\citeauthoryear{{Negueruela} \& {Coe}}{{Negueruela} \&
  {Coe}}{2002}]{negueruela2002}
{Negueruela} I.,  {Coe} M.~J.,  2002, \mn@doi [\aap]
  {10.1051/0004-6361:20020139}, \href
  {https://ui.adsabs.harvard.edu/abs/2002A&A...385..517N} {385, 517}

\bibitem[\protect\citeauthoryear{{Pacucci}, {Mesinger}, {Mineo}  \&
  {Ferrara}}{{Pacucci} et~al.}{2014}]{pacucci2014}
{Pacucci} F.,  {Mesinger} A.,  {Mineo} S.,   {Ferrara} A.,  2014, \mn@doi
  [\mnras] {10.1093/mnras/stu1240}, \href
  {http://adsabs.harvard.edu/abs/2014MNRAS.443..678P} {443, 678}

\bibitem[\protect\citeauthoryear{{Parsons} et~al.,}{{Parsons}
  et~al.}{2014}]{parsons13}
{Parsons} A.~R.,  et~al., 2014, \mn@doi [\apj] {10.1088/0004-637X/788/2/106},
  \href {http://adsabs.harvard.edu/abs/2014ApJ...788..106P} {788, 106}

\bibitem[\protect\citeauthoryear{{Patil} et~al.,}{{Patil}
  et~al.}{2017}]{2017ApJ...838...65P}
{Patil} A.~H.,  et~al., 2017, \mn@doi [\apj] {10.3847/1538-4357/aa63e7}, \href
  {http://adsabs.harvard.edu/abs/2017ApJ...838...65P} {838, 65}

\bibitem[\protect\citeauthoryear{{Patra}, {Subrahmanyan}, {Sethi}, {Udaya
  Shankar}  \& {Raghunathan}}{{Patra} et~al.}{2015}]{2015ApJ...801..138P}
{Patra} N.,  {Subrahmanyan} R.,  {Sethi} S.,  {Udaya Shankar} N.,
  {Raghunathan} A.,  2015, \mn@doi [\apj] {10.1088/0004-637X/801/2/138}, \href
  {http://adsabs.harvard.edu/abs/2015ApJ...801..138P} {801, 138}

\bibitem[\protect\citeauthoryear{{Paul}}{{Paul}}{2017}]{paul2017}
{Paul} B.,  2017, \mn@doi [Journal of Astrophysics and Astronomy]
  {10.1007/s12036-017-9475-4}, \href
  {https://ui.adsabs.harvard.edu/#abs/2017JApA...38...39P} {38, 39}

\bibitem[\protect\citeauthoryear{{Planck Collaboration} et~al.,}{{Planck
  Collaboration} et~al.}{2014}]{Planck2013}
{Planck Collaboration} et~al., 2014, \mn@doi [\aap]
  {10.1051/0004-6361/201321591}, \href
  {http://adsabs.harvard.edu/abs/2014A%26A...571A..16P} {571, A16}

\bibitem[\protect\citeauthoryear{{Planck Collaboration} et~al.,}{{Planck
  Collaboration} et~al.}{2016}]{2016A&A...596A.108P}
{Planck Collaboration} et~al., 2016, \mn@doi [\aap]
  {10.1051/0004-6361/201628897}, \href
  {http://adsabs.harvard.edu/abs/2016A%26A...596A.108P} {596, A108}

\bibitem[\protect\citeauthoryear{{Power}, {Wynn}, {Combet}  \&
  {Wilkinson}}{{Power} et~al.}{2009}]{power2009}
{Power} C.,  {Wynn} G.~A.,  {Combet} C.,   {Wilkinson} M.~I.,  2009, \mn@doi
  [\mnras] {10.1111/j.1365-2966.2009.14628.x}, \href
  {http://adsabs.harvard.edu/abs/2009MNRAS.395.1146P} {395, 1146}

\bibitem[\protect\citeauthoryear{{Power}, {James}, {Combet}  \& {Wynn}}{{Power}
  et~al.}{2013}]{power2013}
{Power} C.,  {James} G.,  {Combet} C.,   {Wynn} G.,  2013, \mn@doi [\apj]
  {10.1088/0004-637X/764/1/76}, \href
  {http://adsabs.harvard.edu/abs/2013ApJ...764...76P} {764, 76}

\bibitem[\protect\citeauthoryear{{Pradhan}, {Bozzo}  \& {Paul}}{{Pradhan}
  et~al.}{2018}]{pradhan2018}
{Pradhan} P.,  {Bozzo} E.,   {Paul} B.,  2018, \mn@doi [\aap]
  {10.1051/0004-6361/201731487}, \href
  {https://ui.adsabs.harvard.edu/\#abs/2018A&A...610A..50P} {610, A50}

\bibitem[\protect\citeauthoryear{{Price} et~al.,}{{Price}
  et~al.}{2018}]{price2018}
{Price} D.~C.,  et~al., 2018, \mn@doi [\mnras] {10.1093/mnras/sty1244}, \href
  {https://ui.adsabs.harvard.edu/\#abs/2018MNRAS.478.4193P} {478, 4193}

\bibitem[\protect\citeauthoryear{{Pritchard} \& {Furlanetto}}{{Pritchard} \&
  {Furlanetto}}{2007}]{pritchard2007}
{Pritchard} J.~R.,  {Furlanetto} S.~R.,  2007, \mn@doi [\mnras]
  {10.1111/j.1365-2966.2007.11519.x}, \href
  {http://adsabs.harvard.edu/abs/2007MNRAS.376.1680P} {376, 1680}

\bibitem[\protect\citeauthoryear{{Reig}}{{Reig}}{2011}]{reig2011}
{Reig} P.,  2011, \mn@doi [\apss] {10.1007/s10509-010-0575-8}, \href
  {https://ui.adsabs.harvard.edu/abs/2011Ap&SS.332....1R} {332, 1}

\bibitem[\protect\citeauthoryear{{Remillard} \& {McClintock}}{{Remillard} \&
  {McClintock}}{2006}]{remillard2006}
{Remillard} R.~A.,  {McClintock} J.~E.,  2006, \mn@doi [\araa]
  {10.1146/annurev.astro.44.051905.092532}, \href
  {http://adsabs.harvard.edu/abs/2006ARA%26A..44...49R} {44, 49}

\bibitem[\protect\citeauthoryear{{Rodes-Roca}, {Mihara}, {Nakahira},
  {Torrej{\'o}n}, {Gim{\'e}nez-Garc{\'{\i}}a}  \& {Bernab{\'e}u}}{{Rodes-Roca}
  et~al.}{2015}]{roca2015}
{Rodes-Roca} J.~J.,  {Mihara} T.,  {Nakahira} S.,  {Torrej{\'o}n} J.~M.,
  {Gim{\'e}nez-Garc{\'{\i}}a} {\'A}.,   {Bernab{\'e}u} G.,  2015, \mn@doi
  [\aap] {10.1051/0004-6361/201425323}, \href
  {http://adsabs.harvard.edu/abs/2015A%26A...580A.140R} {580, A140}

\bibitem[\protect\citeauthoryear{{Ross}, {Dixon}, {Ghara}, {Iliev}  \&
  {Mellema}}{{Ross} et~al.}{2018}]{2018arXiv180803287R}
{Ross} H.~E.,  {Dixon} K.~L.,  {Ghara} R.,  {Iliev} I.~T.,   {Mellema} G.,
  2018, arXiv e-prints, \href
  {https://ui.adsabs.harvard.edu/\#abs/2018arXiv180803287R} {p.
  arXiv:1808.03287}

\bibitem[\protect\citeauthoryear{{Salvaterra}, {Ferrara}  \&
  {Dayal}}{{Salvaterra} et~al.}{2011}]{2011MNRAS.414..847S}
{Salvaterra} R.,  {Ferrara} A.,   {Dayal} P.,  2011, \mn@doi [\mnras]
  {10.1111/j.1365-2966.2010.18155.x}, \href
  {https://ui.adsabs.harvard.edu/abs/2011MNRAS.414..847S} {414, 847}

\bibitem[\protect\citeauthoryear{{Singh} et~al.,}{{Singh}
  et~al.}{2017}]{singh2017}
{Singh} S.,  et~al., 2017, \mn@doi [\apj] {10.3847/2041-8213/aa831b}, \href
  {https://ui.adsabs.harvard.edu/\#abs/2017ApJ...845L..12S} {845, L12}

\bibitem[\protect\citeauthoryear{{Sokolowski} et~al.,}{{Sokolowski}
  et~al.}{2015}]{2015PASA...32....4S}
{Sokolowski} M.,  et~al., 2015, \mn@doi [\pasa] {10.1017/pasa.2015.3}, \href
  {http://adsabs.harvard.edu/abs/2015PASA...32....4S} {32, e004}

\bibitem[\protect\citeauthoryear{{Sugizaki} et~al.,}{{Sugizaki}
  et~al.}{2011}]{sugizaki2011}
{Sugizaki} M.,  et~al., 2011, \pasj, \href
  {http://adsabs.harvard.edu/abs/2011PASJ...63S.635S} {63, 635}

\bibitem[\protect\citeauthoryear{{Tsygankov}, {Wijnands}, {Lutovinov},
  {Degenaar}  \& {Poutanen}}{{Tsygankov} et~al.}{2017}]{tsygankov2017}
{Tsygankov} S.~S.,  {Wijnands} R.,  {Lutovinov} A.~A.,  {Degenaar} N.,
  {Poutanen} J.,  2017, \mn@doi [\mnras] {10.1093/mnras/stx1255}, \href
  {https://ui.adsabs.harvard.edu/\#abs/2017MNRAS.470..126T} {470, 126}

\bibitem[\protect\citeauthoryear{{Voytek}, {Natarajan}, {J{\'a}uregui
  Garc{\'{\i}}a}, {Peterson}  \& {L{\'o}pez-Cruz}}{{Voytek}
  et~al.}{2014}]{2014ApJ...782L...9V}
{Voytek} T.~C.,  {Natarajan} A.,  {J{\'a}uregui Garc{\'{\i}}a} J.~M.,
  {Peterson} J.~B.,   {L{\'o}pez-Cruz} O.,  2014, \mn@doi [\apjl]
  {10.1088/2041-8205/782/1/L9}, \href
  {http://adsabs.harvard.edu/abs/2014ApJ...782L...9V} {782, L9}

\bibitem[\protect\citeauthoryear{{Vulic} et~al.,}{{Vulic}
  et~al.}{2018}]{vulic2018}
{Vulic} N.,  et~al., 2018, \mn@doi [\apj] {10.3847/1538-4357/aad500}, \href
  {https://ui.adsabs.harvard.edu/#abs/2018ApJ...864..150V} {864, 150}

\bibitem[\protect\citeauthoryear{{Wen}, {Levine}, {Corbet}  \& {Bradt}}{{Wen}
  et~al.}{2006}]{wen2006}
{Wen} L.,  {Levine} A.~M.,  {Corbet} R.~H.~D.,   {Bradt} H.~V.,  2006, \mn@doi
  [\apjs] {10.1086/500648}, \href
  {http://adsabs.harvard.edu/abs/2006ApJS..163..372W} {163, 372}

\bibitem[\protect\citeauthoryear{{White}, {Swank}  \& {Holt}}{{White}
  et~al.}{1983}]{white1983}
{White} N.~E.,  {Swank} J.~H.,   {Holt} S.~S.,  1983, \mn@doi [\apj]
  {10.1086/161162}, \href
  {https://ui.adsabs.harvard.edu/#abs/1983ApJ...270..711W} {270, 711}

\bibitem[\protect\citeauthoryear{{Wise}, {Turk}, {Norman}  \& {Abel}}{{Wise}
  et~al.}{2012}]{2012ApJ...745...50W}
{Wise} J.~H.,  {Turk} M.~J.,  {Norman} M.~L.,   {Abel} T.,  2012, \mn@doi
  [\apj] {10.1088/0004-637X/745/1/50}, \href
  {https://ui.adsabs.harvard.edu/abs/2012ApJ...745...50W} {745, 50}

\bibitem[\protect\citeauthoryear{{Wouthuysen}}{{Wouthuysen}}{1952}]{wouth52}
{Wouthuysen} S.~A.,  1952, \mn@doi [\aj] {10.1086/106661}, \href
  {http://adsabs.harvard.edu/abs/1952AJ.....57R..31W} {57, 31}

\bibitem[\protect\citeauthoryear{{Yukita} et~al.,}{{Yukita}
  et~al.}{2016}]{yukita2016}
{Yukita} M.,  et~al., 2016, \mn@doi [\apj] {10.3847/0004-637X/824/2/107}, \href
  {http://adsabs.harvard.edu/abs/2016ApJ...824..107Y} {824, 107}

\bibitem[\protect\citeauthoryear{{Zdziarski}, {Johnson}  \&
  {Magdziarz}}{{Zdziarski} et~al.}{1996}]{zdziarski1996}
{Zdziarski} A.~A.,  {Johnson} W.~N.,   {Magdziarz} P.,  1996, \mn@doi [\mnras]
  {10.1093/mnras/283.1.193}, \href
  {http://adsabs.harvard.edu/abs/1996MNRAS.283..193Z} {283, 193}

\bibitem[\protect\citeauthoryear{{Zhang}, {Gilfanov}  \& {Bogd{\'a}n}}{{Zhang}
  et~al.}{2012}]{zhang2012}
{Zhang} Z.,  {Gilfanov} M.,   {Bogd{\'a}n} {\'A}.,  2012, \mn@doi [\aap]
  {10.1051/0004-6361/201219015}, \href
  {http://adsabs.harvard.edu/abs/2012A%26A...546A..36Z} {546, A36}

\makeatother
\end{thebibliography}
\label{lastpage}               
\end{document}